\documentclass[prc,onecolumn]{revtex4}
\usepackage{amsmath,latexsym,amssymb}
\usepackage[dvips]{graphicx}
\usepackage{epsfig}

\def\vec#1{\mathchoice
 {\mbox{\boldmath $\displaystyle#1$}}
 {\mbox{\boldmath $\textstyle#1$}}
 {\mbox{\boldmath $\scriptstyle#1$}}
 {\mbox{\boldmath $\scriptstyle#1$}}}

\begin{document}

\title{Collective excitations and energy-weighted sum rules in relativistic
random phase approximation with vacuum polarization}

\author{
A. Haga$^{1,2}$\footnote{Electronic address:
haga@rcnp.osaka-u.ac.jp}, H. Toki$^1$\footnote{Electronic
address: toki@rcnp.osaka-u.ac.jp}, S.
Tamenaga$^1$\footnote{Electronic address:
stame@rcnp.osaka-u.ac.jp}, and Y. Horikawa$^3$\footnote{Electronic
address: horikawa@sakura.juntendo.ac.jp} }
\affiliation{ $^1$Research Center for Nuclear Physics (RCNP),
Osaka University, Ibaraki, Osaka 567-0047, Japan\\
$^2$Department
of Engineering Physics, Electronics, and Mechanics,
Nagoya Institute of Technology, Gokiso, Nagoya 466-8555, Japan\\
$^3$Department of Physics, Juntendo University, Inba-gun, Chiba
270-1695, Japan
}

\setlength{\baselineskip}{0.3in}


\begin{abstract}
\setlength{\baselineskip}{0.3in}

The isoscalar and isovector collective multipole excitations in stable nuclei are 
studied in the framework of relativistic random-phase approximation with the vacuum polarization
arising from the nucleon-antinucleon field.
A fully self-consistent calculation which guarantees the decoupling of the spurious state and 
the conservation of the multipole-transition current is carried out by using the derivative-expansion method for the description of the vacuum contribution.
A remarkable effect of the inclusion of the vacuum polarization is the increase of the effective mass, $m^*/m_N \sim 0.8$; 
for all multipole modes, the energy-weighted sum rule values with the vacuum polarization are smaller than 
those of the relativistic model without the vacuum polarization, 
which typically has the effective mass $m^*/m_N \sim 0.6$.
Also, the present model can give an excellent agreement with experimental data on the excitation 
energy, in particular, for the isoscalar quadrupole resonances in which it was previously reported that
the calculated energies in the relativistic model are about 1-2 MeV above the experimental values. 
It is shown, further, that the change of the shell structure due to the inclusion of the vacuum polarization
plays an important role in the improvement of the discrepancies seen in the dipole compression modes. 
On the other hand, the isoscalar monopole resonance has a similar peak
whether or not the vacuum polarization is taken into account, if the compression modulus is kept the same in the analysis.

PACS number(s): 21.10.-k, 21.60.Jz, 13.75.Cs, 23.20.-g
\end{abstract}

\maketitle

\section{Introduction}

The giant resonance of nuclei is of great interest
because its energy can be directly related to the compressibility of nuclear matter\cite{BL80,ST82},
which places important constraints on the theoretical description of nuclear properties,
supernova explosions, neutron stars, heavy ion collisions, and so on.
The reliability of the model employed for the giant-resonance analyses
may be thus confirmed by the comparison with the experimental energy-weighted sum rule (EWSR) distributions.
On the other hand, only the ratio of the EWSR between different nuclei, rather than the EWSR strength,
was available because the sum rule strength was dependent on model assumptions, as pointed out in Ref.~\cite{YO97}.
A knowledge of the magnitude of the EWSR for the electromagnetic response in addition to its distribution, however, 
will impose a further constraint with regard to the nuclear model, and offer valuable information for 
the study of a two-photon exchange process through the nuclear particle-hole excitations such as 
the dispersion correction to the elastic electron scattering\cite{FR74} and nuclear-polarization correction to the muonic levels\cite{RI82}, 
which are approximately proportional to the EWSR strength.

The comparison between the experimental data and the theoretical analysis on the nuclear excitations
provides a critical test of the quantum hadrodynamics (QHD) in a many-body system\cite{WA74}.
A standard tool for the theoretical description of nuclear excitations is the random-phase approximation (RPA).
In particular, many studies have been done by using relativistic RPA (RRPA) with the relativistic mean-field (RMF) basis
\cite{FU85,KUSU85,IC87,WE87,HOPI88,FU89,MC89,DA90,MA97,MA01,CO98,PI00,RI01,VE00,GR03}.
On the basis of the works of the last decade, it has been realized that the self-consistent method with nonlinear effective Lagrangians 
is essential for a quantitative description of excited states.
In this context, it is very interesting that the negative-energy nucleon state in the mean-field basis seriously contributes 
to excitation energies\cite{RI01}, the current conservation of the transition 
densities\cite{MC89,DA90,PI00}, and the decoupling of the center-of-mass motion\cite{DA90,PI00}. 
Furthermore, it was found that the negative-energy RRPA eigenstates, 
generated from the RRPA equation with the fully consistent
basis including negative-energy RMF states, have a significant role in 
gauge invariance of the response to an electromagnetic field\cite{HA041}.

The success of the relativistic nuclear model is mainly ascribed to the small effective mass, $m^*/m_N \sim 0.6$,
which yields the appropriate single-particle splittings in a nuclear ground state.
The fact that the centroid (or peak) energies and the distributions of the nuclear excitations in the isoscalar monopole mode
and the isovector dipole mode are well reproduced\cite{RI01} seems to support the validity of the small effective mass.
However, the physical quantities which are more sensitive to the variety of the nucleon mass are
the strength of the response function and the centroid energy of the giant-resonance mode which the shell structure around the Fermi surface affects significantly. 
From this viewpoint, an interesting fact in the recent study of Ref.~\cite{NI05} 
is that the $\beta$-decay lifetimes require an enhanced effective mass to reproduce the data.

The small effective mass is caused by ignoring negative-energy nucleons (or antinucleons).
If the Dirac sea filled by negative-energy nucleons is duly taken into account, then the effective mass converges at an enhanced value
naturally because the Lorentz scalar field is strongly suppressed due to the feedback effect from antinucleon states\cite{HA042}.
Inclusion of the Dirac sea, however, raises a problem in fitting the single-particle energies, requiring a new mechanism to supply the spin-orbit force.
A simplest solution would be to add the tensor-coupling term for the meson-nucleon interaction to the Lagrangian\cite{SU942,MAO03}.
In this work we employ a new extended relativistic model with tensor coupling for the vector meson incorporating the Dirac sea effects of nucleons.
We emphasize that the self-consistency is kept throughout the RHA + RRPA calculation, as verified by observing 
the spurious state in the isoscalar dipole mode and the current conservation of the transition densities\cite{HA05}.
We present a systematic study of the giant monopole, dipole, and quadrupole resonances
for the different effective mass in the fully consistent RRPA framework.

This paper is organized as follows. In Sec.~II
we introduce the effective Lagrangian with renormalized negative-energy
nucleon contribution by using the derivative-expansion method.
With this approach, the contribution from negative-energy nucleons to the effective action can be expressed by the functional of meson fields.
Such a action gives the meson propagators dressed 
by the vacuum polarization of nucleon-antinucleon pairs automatically.
The details of the computational procedure for the self-consistent RHA + RRPA calculation 
with the vacuum correction are described in Sec.~III.
The vacuum effect to the decoupling of the center-of-mass motion and the way to conserve the RPA transition densities 
in the model, including the tensor-coupling interaction, are also shown here.
In Sec.~IV, the numerical results of the present calculation for the Coulomb response 
function in the spherical nuclei are compared with those of the TM1 and NL3 parameter sets.
In particular, the giant monopole, dipole, and quadrupole resonances of the $^{90}$Zr, $^{116}$Sn, and $^{208}$Pb
as well as their energy-weighted sum rules are discussed in detail.
Finally, we give a summary of our calculations and an outlook of future applications in Sec.~V.

\section{formalism}

Within a relativistic framework, it is inevitable that the negative-energy nucleon states 
contribute to the nuclear structure as a vacuum fluctuation. 
Recent investigations of the RRPA in the QHD model, however, have neglected the actual antinucleon degrees of freedom;
the basis set used for the RRPA calculation is prepared by the RMF theory wherein 
only positive-energy nucleons are taken into account, so that 
the nuclear-polarization tensor in the RRPA equation is described without the vacuum polarization of 
the nucleon-antinucleon field.
On the other hand, the rigorous calculation of basis set under the full one-nucleon-loop contribution,
referred to as the relativistic Hartree approach (RHA), has been developed by using the Green function method\cite{HA042,BL90,ST97}. 
With this technique, however, the estimation of the vacuum-polarization effect is too hard even for the ground state of 
the light nucleus to extend it to the RRPA formalism including the particle-hole correlations, especially for heavy nuclei.
Fortunately, it was found in Ref.~\cite{HA042} that the lowest-order contribution of the derivative expansion 
from the one-baryon-loop correction to the effective action is in excellent agreement with the rigorous result.
The derivative-expansion method describes the vacuum contribution by the additional new terms of meson fields
in the effective Lagrangian\cite{JA74}. 
As a result, it makes it easy to perform the RRPA calculation with vacuum polarization because the meson propagators 
are constructed by including the vacuum part of the nuclear response as well as self-interaction terms\cite{HA05}. 

A remaining problem in nuclear models with the vacuum-polarization effect
was that the magnitude of the spin-orbit splittings in the shell structure of the mean field
is insufficient in comparison with the observed data because the vacuum-polarization effect decreases 
the spin-orbit potential of nucleons. 
Reasonable results for the ground-state properties of the binding energy and the charge radius 
as well as the spin-orbit splittings were given by Mao\cite{MAO03}, who has investigated the single-particle levels of nucleons and antinucleons 
by incorporating the effect of the tensor-coupling meson-nucleon interaction.
As far as the spin-orbit splittings are concerned, 
the tensor coupling of meson-nucleon interaction, which directly generates the spin-orbit force in the RHA calculation, 
is considerably effective for the simple extension of the model in comparison with other candidates 
such as the surface pion condensation\cite{OG04} and the Fock contribution of the pion exchange.  
Hence, in the present paper we shall also incorporate the tensor-coupling interaction for the $\omega$ meson.
Although the model is now non-renormalizable and the divergence should be removed by taking an infinite number of couplings into account,
we assume that such a vacuum contribution from the tensor-coupling interaction is negligible.
The following Lagrangian density is thus employed for the present model;
\begin{align}
\mathcal{L}^{ren}_{(1)}=&\bar{\psi}^+ 
\left[i\gamma_\mu
\partial^\mu-m_N+g_\sigma\sigma-g_\omega\gamma^\mu\omega_\mu
-\frac{f_\omega}{4m_N}\sigma^{\mu\nu}(\partial_{\mu}\omega_{\nu}-\partial_{\nu}\omega_{\mu})
-g_\rho\gamma^\mu \frac{\vec{\tau}}{2} \cdot \vec{\rho}_\mu
-e{\cal Q}\gamma^\mu A_\mu
\right]
\psi^+
\nonumber\\
&
+\frac{1}{2}(\partial_\mu\sigma)^2-\frac{1}{2}m_\sigma^2\sigma^2
-U_\sigma(\sigma)
\nonumber\\
&
-\frac{1}{4}(\partial_{\mu}\omega_{\nu}-\partial_{\nu}\omega_{\mu})^2
+\frac{1}{2}m_\omega^2(\omega_{\mu})^2
+U_\omega(\omega_\mu)
\nonumber\\
&
-\frac{1}{4}(\partial_{\mu}\vec{\rho}_{\nu}-\partial_{\nu}\vec{\rho}_{\mu})^2
+\frac{1}{2}m_\rho^2(\vec{\rho}_{\mu})^2
\nonumber\\
&
-\frac{1}{4}(\partial_{\mu} A_{\nu}-\partial_{\nu} A_{\mu})^2
\nonumber\\
&
-V_F(\sigma)+\frac{1}{2}Z_F^\sigma(\sigma)(\partial_\mu \sigma)^2
+\frac{1}{4}Z_F^\omega(\sigma)
(\partial_{\mu}\omega_{\nu}-\partial_{\nu}\omega_{\mu})^2
+\frac{1}{4}Z_F^A(\sigma)
(\partial_{\mu}A_{\nu}-\partial_{\nu}A_{\mu})^2,
\label{LDE}
\end{align}
where $\psi$ is the Dirac spinor of the nucleon, and $\sigma$, $\omega_\mu$, $\vec{\rho}_\mu$,
and $A_\mu$ represent the isoscalar-scalar meson, isoscalar-vector meson, isovector-vector meson fields, and the electromagnetic field, respectively.
Here $\sigma_{\mu\nu}=(i/2)[\gamma_\mu,\gamma_\nu]$ is the spin tensor, ${\cal Q}=(1+\tau_3)/2$ projects the proton, and
$U_\sigma(\sigma)=\frac{1}{3!}g_{2}\sigma^3+\frac{1}{4!}g_{3}\sigma^4$ ($U_\omega(\omega_\mu)=\frac{1}{4}c_3(\omega_\mu \omega^\mu)^2$)
denotes the self interaction of isoscalar-scalar (-vector) meson.
The superscript $+$ in the nucleon-field operators in Eq.~(\ref{LDE}) means that
only the positive-energy states are to be treated explicitly.
The functions $V_F(\sigma)$, $Z_F^\sigma(\sigma)$, 
$Z_F^\omega(\sigma)$, and $Z_F^A(\sigma)$ which describe the vacuum effect of nucleons
are obtained by well-known prescriptions
\cite{AI84,CH85,OCH85} 
and read as
\begin{align}
V_F(\sigma)&=-\frac{1}{4\pi}
\left[
(m_N-g_\sigma\sigma)^4 \ln\left( 1-\frac{g_\sigma\sigma}{m_N} \right)
+m_N^3 g_\sigma\sigma
-\frac{7}{2}m_N^2 g_\sigma^2\sigma^2
+\frac{13}{3}m_Ng_\sigma^3\sigma^3
-\frac{25}{12}g_\sigma^4\sigma^4
\right],
\\
Z_F^\sigma(\sigma)&=-\frac{g_\sigma^2}{2\pi^2} \ln\left( 1-\frac{g_\sigma\sigma}{m_N} \right),
\\
Z_F^\omega(\sigma)&=\frac{g_\omega^2}{3\pi^2} \ln\left( 1-\frac{g_\sigma\sigma}{m_N} \right),
\\
Z_F^A(\sigma)&=\frac{e^2}{6\pi^2} \ln\left( 1-\frac{g_\sigma\sigma}{m_N} \right).
\label{FCOE}
\end{align}
The paths of classical meson and photon fields are determined under the condition that the contribution 
from the first order of the quantum fluctuations vanishes in the action.  
Assuming the stationary and spherical system, the classical fields 
$\bar{\sigma}$, $\bar{\omega}_0$, $\bar{\rho}_{0,3}$, and $\bar{A}_0$
satisfy the following coupled equations:
\begin{align}
\label{KGS}
(-\nabla^2+m_\sigma^2)\bar{\sigma}&=
g_\sigma\langle \bar{\psi}^+\psi^+ \rangle
-U^\prime_\sigma (\bar{\sigma})-V_F^{\prime}(\bar{\sigma})
-\frac{1}{2}Z_F^{\sigma \prime}(\bar{\sigma})(\nabla \bar{\sigma})^2
+\nabla \cdot (Z_F^\sigma(\bar{\sigma}) \nabla \bar{\sigma})
\nonumber\\
&-\frac{1}{4}{Z_F^{\omega}}^{\prime}(\bar{\sigma})(\nabla \bar{\omega}_0)^2
-\frac{1}{4}{Z_F^{A}}^{\prime}(\bar{\sigma})(\nabla \bar{A}_0)^2,
\\
\label{KGW}
(-\nabla^2+m_\omega^2)\bar{\omega}_0&=g_\omega\langle \bar{\psi}^+\gamma^0\psi^+ \rangle
-U^\prime_\omega(\bar{\omega}_0)
+\frac{f_\omega}{2m_N} \nabla\cdot \langle \bar{\psi}^+\vec{\sigma}\psi^+ \rangle
-\nabla \cdot (Z_F^\omega(\bar{\sigma}) \nabla \bar{\omega}_0),
\\
\label{KGR}
(-\nabla^2+m_\rho^2)\bar{\rho}_{0,3}&=\frac{1}{2}g_\rho\langle \bar{\psi}^+\tau_3\gamma^0\psi^+ \rangle,
\\
\label{KGA}
-\nabla^2 \bar{A}_0&=e\langle \bar{\psi}^+{\cal Q}\gamma^0\psi^+ \rangle
-\nabla \cdot (Z_F^A(\bar{\sigma}) \nabla \bar{A}_0),
\end{align}
The contributions from negative-energy nucleon states to source terms are contained in $V_F$ and $Z_F$.
In a mean-field approximation, nucleon states satisfy a Dirac equation under 
the classical fields $\bar{\sigma}$, $\bar{\omega}_0$, $\bar{\rho}_{0,3}$, and $\bar{A}_0$:
\begin{align}
\left[ -i\gamma^0\vec{\gamma}\cdot (\vec{\nabla}+\frac{f_\omega}{2m_N}\gamma^0\vec{\nabla}\bar{\omega}_0)
+\gamma^0(m_N-g_\sigma\bar{\sigma})
+(g_\omega\bar{\omega}_0
+g_\rho\frac{\tau_3}{2}\bar{\rho}_{0,3}
+e{\cal Q} \bar{A}_0)
\right]
\psi_{\alpha}=
\omega_\alpha \psi_{\alpha},
\label{Dirac}
\end{align}
for which there are solutions with negative energies as well as ordinary positive energies.
The RHA basis consists of the positive- and negative-energy eigenstates obtained 
by solving the Dirac equation (\ref{Dirac}) and the Laplace equations (\ref{KGS})-(\ref{KGA}) iteratively,
until convergence is reached.

\begin{figure}[h]
\centering
\includegraphics[width=7.0cm,clip]{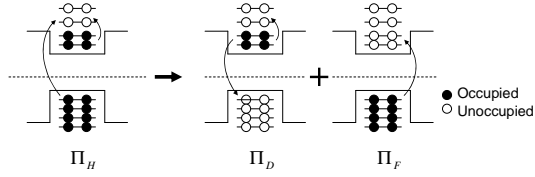}
\caption{\label{fig1}
Schematic representation of the polarization tensor:
$\Pi_D$ represents a transition from the Fermi sea to unoccupied positive- and negative-energy states,
while $\Pi_F$ represents a transition from the Dirac sea to unoccupied positive-energy states.
$\Pi_F$ is divergent without the counterterms.
In the present work, its physical contribution is evaluated by using the derivative-expansion method.}
\end{figure}

Since the vacuum-polarization effect is taken into account in the RHA basis, the RPA diagram also has
to involve the vacuum contribution to retain the consistency.
This is achieved by solving the following Bete-Salpeter equation with the full-unperturbed polarization tensor $\Pi_H$:
\begin{align}
\Pi_{RPA}(\Gamma^a,\Gamma^b;\vec{p},\vec{q};\omega)=&\Pi_H(\Gamma^a,\Gamma^b;\vec{p},\vec{q};\omega)
\nonumber\\
&+\sum_{i} 
\int \frac{d\vec{k}_1}{(2\pi)^3} \frac{d\vec{k}_2}{(2\pi)^3}
\Pi_H(\Gamma^a,\Gamma^i;\vec{p},\vec{k}_1;\omega)
D_{i}^0(\vec{k}_1,\vec{k}_2;\omega) \Pi_{RPA}(\Gamma^i,\Gamma^b;\vec{k}_2,\vec{q};\omega),
\label{ERPAeq}
\end{align}
where the summation is over the meson fields, and $D_i^0$ denotes the meson propagators with coupling constants: 
\begin{align}
D_\sigma^0(\vec{k}_1,\vec{k}_2;\omega)= & -\frac{g_\sigma^2}{\omega^2-\vec{k}_1^2-m_\sigma^2}(2\pi)^3\delta(\vec{k}_1-\vec{k}_2)
\nonumber \\
& -\frac{g_\sigma^2}{\omega^2-\vec{k}_1^2-m_\sigma^2}
\int \frac{d\vec{k}_3}{(2\pi)^3} U^{\prime\prime}_\sigma(\bar{\sigma})(\vec{k}_1,\vec{k}_3) D_\sigma^0(\vec{k}_3,\vec{k}_2;\omega),
\\
D_\omega^0(\vec{k}_1,\vec{k}_2;\omega)= &\frac{g_\omega^2}{\omega^2-\vec{k}_1^2-m_\omega^2}(2\pi)^3\delta(\vec{k}_1-\vec{k}_2)
\nonumber \\
& +\frac{g_\omega^2}{\omega^2-\vec{k}_1^2-m_\omega^2}
\int \frac{d\vec{k}_3}{(2\pi)^3} U^{\prime\prime}_\omega(\bar{\omega})(\vec{k}_1,\vec{k}_3) D_\omega^0(\vec{k}_3,\vec{k}_2;\omega),
\\
D_\rho^0(\vec{k}_1,\vec{k}_2;\omega)= &\frac{g_\rho^2}{\omega^2-\vec{k}_1^2-m_\rho^2}(2\pi)^3\delta(\vec{k}_1-\vec{k}_2),
\\
D_A^0(\vec{k}_1,\vec{k}_2;\omega)= & \frac{e^2}{\omega^2-\vec{k}_1^2}(2\pi)^3\delta(\vec{k}_1-\vec{k}_2),
\end{align}
where the $\sigma$- and $\omega$-meson propagators have to be constructed with the self-interaction term\cite{MA97}.
$\Gamma$'s in Eq.~(\ref{ERPAeq}) are the $4 \times 4$ matrices which denote the vertex couplings,
\begin{align}
\Gamma \in 
{1,\gamma_\mu+\frac{f_\omega}{2m_N}\overleftrightarrow{\partial}^\nu\sigma_{\mu\nu},\frac{\vec{\tau}}{2}\gamma_\mu,{\cal Q}\gamma_\mu },
\end{align}
to the $\sigma$, $\omega$, $\rho$ mesons, and photon, respectively.
The concrete expression of $\Pi_H$ is given by,
\begin{align}
\label{PiH}
\Pi_H (\Gamma^a,\Gamma^b;\vec{x},\vec{y};\omega) &= 
-i\int \frac{d\omega'}{2\pi} {\rm Tr}
\Gamma^a G_H (\vec{x},\vec{y};\omega+\omega') \Gamma^b G_H (\vec{y},\vec{x};\omega')
\nonumber\\
&=\sum_{\alpha\beta} 
\bar{\psi}_\beta(\vec{x})\Gamma^a\psi_\alpha(\vec{x})
\bar{\psi}_\alpha(\vec{y})\Gamma^b\psi_\beta(\vec{y})
(\Pi_D^{\alpha\beta}(\omega)+\Pi_F^{\alpha\beta}(\omega)),
\\
\Pi_D^{\alpha\beta}(\omega) =& \frac{\left(
\theta(\omega_\alpha-\epsilon_F)+\theta(-\omega_\alpha)
\right)
\theta(\epsilon_F-\omega_\beta)\theta(\omega_\beta)}
{\omega-(\omega_\alpha-\omega_\beta)+i\omega_\alpha\eta/2}
-\frac{\left(
\theta(\omega_\beta-\epsilon_F)+\theta(-\omega_\beta)
\right)
\theta(\epsilon_F-\omega_\alpha)\theta(\omega_\alpha)}
{\omega-(\omega_\alpha-\omega_\beta)-i\omega_\beta\eta/2},
%
%
\\
\Pi_F^{\alpha\beta}(\omega) =&
\frac{
\theta(\omega_\alpha)\theta(-\omega_\beta)}
{\omega-(\omega_\alpha-\omega_\beta)+i\eta/2}
-\frac{\theta(-\omega_\alpha)\theta(\omega_\beta)}
{\omega-(\omega_\alpha-\omega_\beta)-i\eta/2},
\end{align}
where $\epsilon_F$ is the Fermi energy.
They are sketched in Fig.~\ref{fig1}, where
$\Pi_D$ is composed of the ordinary particle-hole excitation and 
the Pauli-principle violating excitation between
the positive-energy and the negative-energy nucleon states. 
On the other hand, $\Pi_F$ is composed of the excitation between the positive-energy particle 
and the negative-energy hole as the Fermi sea is unoccupied.
The latter is divergent, and therefore the renormalization procedure is required to 
extract the physical contribution from it.
To our knowledge, $\Pi_F$ has never been calculated explicitly in a finite system and it is still a remaining task.
However, the use of the derivative-expansion method, which renormalizes the divergence in the effective action, 
enable us to evaluate the vacuum contribution to the polarization tensor within this approximation.
As indicated in our previous paper\cite{HA05},
the second-order fluctuation terms of meson fields in the effective action give 
the meson propagators including the vacuum polarization expressed by the functions of meson fields.
For the RPA equation (\ref{ERPAeq}), we shall define the finite unperturbed polarization tensor as
\begin{align}
\label{FULLPI}
\Pi_H=\Pi_D+\tilde{\Pi}_F,
\end{align}
where the vacuum-polarization part $\tilde{\Pi}_F$ 
is given as
\begin{align}
\tilde{\Pi}_F^{\sigma\sigma}(y,x)&= \delta(x-y)
\left[
V_F^{\prime\prime}(\bar{\sigma})
-\frac{1}{2}Z_F^{\sigma \prime\prime}(\bar{\sigma})(\partial_\mu \bar{\sigma})^2
-\frac{1}{2}Z_F^{\omega \prime\prime}(\bar{\sigma})(\partial_\mu \bar{\omega}_0)^2
-\frac{1}{2}Z_F^{A \prime\prime}(\bar{\sigma})(\partial_\mu \bar{A}_0)^2
\right]
\nonumber\\
& \qquad \qquad \qquad 
+ \delta(x-y)[ \partial_\mu \partial^\mu Z^\sigma_F(\bar{\sigma})] 
+ \partial^\mu\left[Z_F^\sigma(\bar{\sigma})\left[\partial_\mu \delta(x-y)
\right]\right],
\label{piss}
\\
\label{piws}
\tilde{\Pi}_F^{\sigma\omega}(y,x)& 
= \partial_\mu[Z_F^{\omega\prime}(\bar{\sigma}) (\partial_\mu \bar{\omega}_0)\delta(x-y)],
\\
\label{pisw}
\tilde{\Pi}_F^{\omega\sigma}(y,x) &
= - Z_F^{\omega\prime}(\bar{\sigma}) (\partial_\mu \bar{\omega}_0)[\partial^\mu \delta(x-y)], 
\\
\label{piww}
\tilde{\Pi}_F^{\omega\omega}(y,x)&= 
\partial^\mu \left[ Z_F^\omega(\bar{\sigma})\left[\partial_\mu \delta(x-y)
\right] \right],
\\
\label{piAs}
\tilde{\Pi}_F^{\sigma A}(y,x)& 
= \partial_\mu[Z_F^{A\prime}(\bar{\sigma}) (\partial_\mu \bar{A}_0)\delta(x-y)],
\\
\label{pisA}
\tilde{\Pi}_F^{A\sigma}(y,x) &
= - Z_F^{A\prime}(\bar{\sigma}) (\partial_\mu \bar{A}_0)[\partial^\mu \delta(x-y)], 
\\
\label{piAA}
\tilde{\Pi}_F^{AA}(y,x)&= 
\partial^\mu \left[ Z_F^A(\bar{\sigma})\left[\partial_\mu \delta(x-y)
\right] \right],
\end{align}
where the superscript in $\tilde{\Pi}_F$ indicates mesons at initial and final vertices.
We have shown in Ref.~\cite{HA042} that 
the leading order of the derivative expansion in RHA calculation reproduces the numerical results of the exact Green function method very well.
It is expected that also in the RPA calculation, it is sufficient to expand up to the leading order of the derivative terms,
as far as the low-momentum transfer region are concerned.
The verification of the self-consistency will provide a check that the vacuum polarizations in the RPA level are 
sufficiently described by Eqs.~(\ref{piss})-(\ref{piAA}).
In the next section, we will show that the spurious state in the isoscalar dipole mode
is decoupled and the transition density is conserved with $\tilde{\Pi}_F$ of Eqs.~(\ref{piss})-(\ref{piAA}).

\section{Numerical detail of vacuum contribution, tensor-coupling effect, and antinucleon solution}

We shall explain here the main points of the calculation of the self-consistent RHA + RRPA, including the vacuum polarization.
First of all, we perform the RHA calculation, which gives self-consistent mean-field potentials 
by making the iteration of the nucleon- and meson-field calculation.  
The parameter sets used in the present work are listed in Table I, where the Fermi momentum, the binding energy,
the effective mass, the compression modulus, and the asymmetry energy in nuclear matter are also shown.
TM1\cite{SU94} and NL3\cite{LA97} are the well-known parameter sets which give good descriptions of nuclear properties and
RHAT1, RHAT2, and RHAT3 are the parameter sets employed for the present model explained in the previous section.
They are determined by firstly surveying the parameters in nuclear matter and choosing the sets
which have the compression modulus $K=270$, $281$, and $325$ MeV 
and then, by fitting to the total binding energies, charge radii, and single-particle energies of the spherical nuclei 
in the RHA calculation, with a fine adjustment of the coupling constants of the $\sigma$ and $\omega$ mesons 
in addition to the coupling constants of the $\rho$ meson and of the $\omega$-nucleon tensor coupling.
As seen in Table II, they reproduce the nuclear ground-state properties as well as TM1 and NL3.

\begin{figure}[h]
\centering
\includegraphics[width=7.0cm,clip]{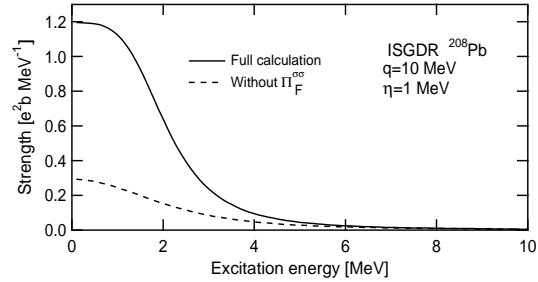}
\caption{\label{Spurious}
ISGDR strength with the parameter set RHAT1 in the momentum transfer $q=10$ MeV.
The spurious state gets shifted to zero energy in the full calculation (the solid line).
If $\tilde{\Pi}_F^{\sigma\sigma}$ is neglected in the RRPA calculation (the dashed line),
the elimination of the spurious state from the real excitation-energy region becomes incomplete.}
\end{figure}

The mean-filed potential of each parameter set is then used to generate the basis set of nucleons.
The RHA basis functions in the RRPA calculation are obtained by solving the single-particle Dirac equation
(\ref{Dirac}) using the method of discretization for the continuum states\cite{DA90}.
For the excited states, we have included the positive-energy states up to $200$ MeV and practically 
all of the negative-energy bound states (up to $-1875$ MeV), which contain a large number of bound states 
in the negative-energy spectrum, since the scalar and vector potentials add in the effective negative-energy potential.
The RRPA equation (\ref{ERPAeq}) is solved directly by the inverse matrix method in the momentum space
which allows the inclusion of as many configurations as required without changing the size of the matrix.
Figure \ref{Spurious} shows the RRPA Coulomb response of the isoscalar dipole resonance in $^{208}$Pb with the parameter set RHAT1.
There, we choose the momentum transfer, $q=10$ MeV, and the width, $\eta=1$ MeV.
The solid curve is the complete result and the dashed one is the result without $\tilde{\Pi}_F^{\sigma\sigma}$ (Eq.~(\ref{piss})), 
ignoring a part of the vacuum polarization, in the RRPA calculation.
We see that the position of the strength peak which is composed of the center-of-mass motion gets shifted to zero energy, 
as expected, in the fully consistent RHA + RRPA calculation.
For the dashed line, on the other hand, we find that the spurious component is removed insufficiently in the real excitation-energy range.
Thus, we conclude that the inclusion of the vacuum polarization defined by Eqs.~(\ref{piss})-(\ref{piAA}) in the RRPA equation
is important for the self-consistency.

\begin{figure}[h]
\centering
\includegraphics[width=7.0cm,clip]{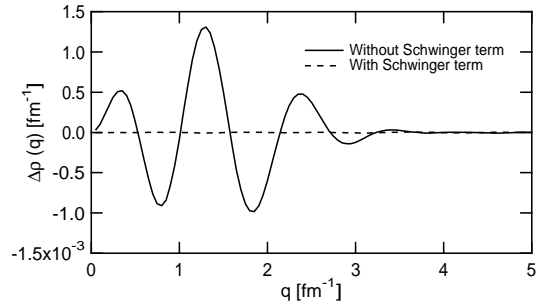}
\caption{\label{Violate}
Violation of the current conservation in the isoscalar quadrupole 3.7-MeV state
of $^{208}$Pb with the parameter set RHAT1.
The violation is quite large if the Schwinger term is neglected in the polarization function (the solid line).
The inclusion of the Schwinger term makes the transition current conserved (the dashed line).}
\end{figure}

Another important aspect for establishing the correctness of the self-consistent RRPA calculations is to check whether 
transition charge density $\rho_\lambda(q)$ and current density $j_{\lambda L}(q)$ 
connecting the ground state and the excited states for different multipolarity
$\lambda$ and $L$, satisfy the conservation law \cite{MC89,DA90,PI00}.
With $\omega_N$ denoting the excitation energy of the nucleus,
the conservation relation is given by,
\begin{align}
\omega_N
\rho_\lambda(q)
&=
-q \sqrt{\frac{\lambda}{2\lambda+1}}
j_{\lambda \lambda-1}(q)
+q
\sqrt{\frac{\lambda+1}{2\lambda+1}}
j_{\lambda \lambda+1}(q).
\label{conservation}
\end{align}
The violation of the current conservation in the transition densities, defined by 
the difference between the left-hand and right-hand sides of Eq.~(\ref{conservation}) (hereafter referred
to as $\Delta \rho$ ) should be very small with a large configuration space. 
However, the RRPA states calculated with the unperturbed-polarization tensor defined by Eq.~(\ref{PiH}) {\it do not satisfy} 
the conservation law as shown by the solid line in Fig.~\ref{Violate}.
This violation is due to the tensor-coupling term for the $\omega$ meson.
In order to make the transition current conserved, the additional term is required in 
the photon-$\omega$ meson polarization tensor (More detailed argument is given in the Appendix A.);
\begin{align}
\label{PiHN}
i\Pi_D ({\cal Q}\gamma^\mu,\gamma^\nu+\frac{f_\omega}{2m_N}\overleftrightarrow{\partial}_\xi\sigma^{\nu\xi};x,y)
\equiv
\langle 0| T[j_A^\mu(x) j_\omega^\nu(y)]|0 \rangle
+\langle 0| [j_A^\mu(x), j_{\omega A}^\nu(y)]_0|0 \rangle,
\end{align}
where $[\cdots]_0$ denotes the equal-time commutation relation, and
\begin{align}
\label{JA}
j_A^\mu  & = \bar{\psi} {\cal Q}\gamma^\mu \psi,
\\
\label{JO}
j_\omega^\nu  & = \bar{\psi} \gamma^\nu \psi+ \frac{f_\omega}{2m_N}\partial_\xi (\bar{\psi}\sigma^{\nu\xi}\psi),
\\
\label{JOA}
j_{\omega A}^\nu & = \frac{f_\omega}{2m_N} \bar{\psi} \sigma^{\nu 0} \psi.
\end{align}
The second term of the right-hand side of Eq.~(\ref{PiHN}) is called the "Schwinger term"\cite{AD68}.
In the present case, the Schwinger term exists in the transverse polarization tensor.
Including this contribution, the $\Delta \rho$ shown by the dashed line
in Fig.~\ref{Violate} becomes negligibly small, that is, the conservation law is achieved.
Since the complete set of the basis is assumed in the proof of the conservation of the RPA transition densities\cite{MC89},
this result also confirms that our truncation in the spectral method 
for the positive and negative energies is reasonable.

Another, and main role of the tensor-coupling term in the present nuclear model is to produce the proper spin-orbit force in the single-particle basis.
For the mean-field calculation, we do not discuss it because Mao has already investigated its effect in detail\cite{MAO03}.
Here, we shall briefly mention the tensor-coupling effect appearing in the RRPA states.
Since the tensor-coupling term restores the spin-orbit force with a large effective mass,
the RRPA states may be influenced by the change of the single-particle shell structure in addition to the residual tensor-coupling interaction.
For the collective modes, however, the calculations with and without the tensor-coupling interaction give the similar result with each other,
if the parameter sets producing the similar compressibility and the similar effective mass are used\cite{HA052}.
A large effect due to the tensor-coupling term appears in the low-lying states, in particular, 
in the higher multipole mode as the quadrupole and octupole excitations, in which the shell structure of the Fermi-energy region concerns directly.
This fact means that the tensor couplings enable us to fit the energy spectrum of the low-lying states phenomenologically 
without changing the giant-resonance structure.

\begin{figure}[h]
\centering
\includegraphics[width=7.0cm,clip]{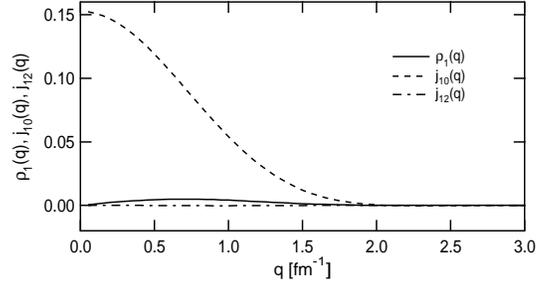}
\caption{\label{negative}
The transition-charge and -current densities in the isovector antinucleon dipole $1507.2$-MeV state
of $^{90}$Zr with the parameter set RHAT1.
The transition-current density $j_{10}(q)$ (the dashed line) is considerably large in comparison with the 
the transition-charge density $\rho_{1}(q)$ (the solid line) and the transition-current density 
$j_{12}(q)$ (the dash-dotted line).}
\end{figure}

In the last part of this section, we shall discuss the RRPA eigenstates with the negative energy.
As discussed in Ref.~\cite{HA041}, the fully-consistent RRPA equation has the solutions for the eigenstates with the negative energy
which represent the blocking effect of the nucleon-antinucleon creation due to the states occupied by the spectator nucleon.
In the present model with the vacuum polarization, a negative-energy solution is interpreted as an antinucleon one and also is composed
of a part of the complete set of the RRPA states together with the ordinary positive-energy solutions.
As an example, the longitudinal (Coulomb) and transverse form factors for the isovector lowest-antinucleon dipole state 
($1507.2$ MeV) with the parameter set RHAT1 is depicted in Fig.~\ref{negative}.
The interesting thing seen in this figure is that the transverse form factor is considerably larger than 
the Coulomb form factor.
It is also seen that the form factors of antinucleon states
have a peak in a very low momentum region than the corresponding transition energy, because the states are bound strongly.
These facts are very similar to those in the RRPA without the vacuum polarization except for the position of the
excitation energy; the potential in the antinucleon sector becomes extremely deep by ignoring the vacuum polarization, so that
the lowest-antinucleon state is located at a lower energy than in the present case.

\section{Results and discussion}

We now discuss the calculated results of the Coulomb response function for collective excitations predicted by the relativistic
nuclear model with the vacuum polarization.
Theoretically, the nucleus cannot be constructed without the large effective mass if the vacuum of nucleons is duly accounted for\cite{HA042,MA99}.
Although the large effective mass changes the shell structure obtained by the models without the vacuum polarization, 
the nuclear ground-state properties come out to be similar for all the three parameter sets used in this work.
For nuclear excitations, however, we can see the vacuum-polarization effect through the effective mass effect
in the EWSR strength, and in the excitation energies of some giant resonances.

\subsection{Energy-weighted sum rules}

\begin{figure}[h]
\centering
\includegraphics[width=7.0cm,clip]{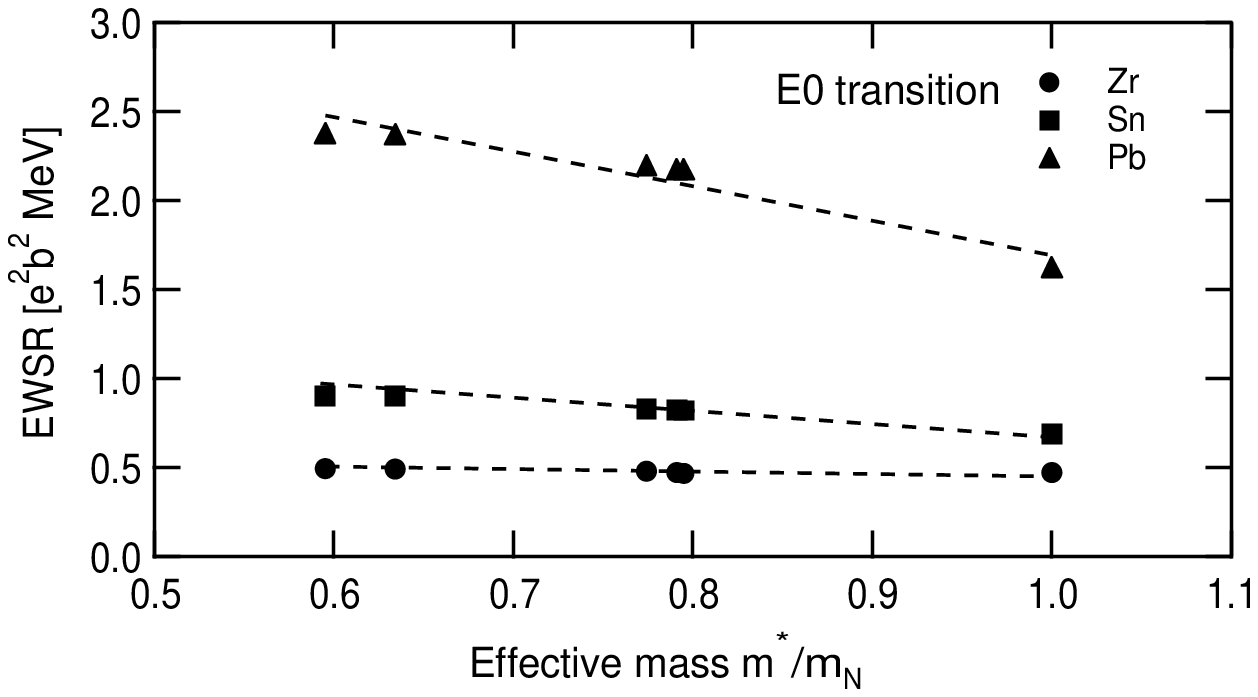}
\includegraphics[width=7.0cm,clip]{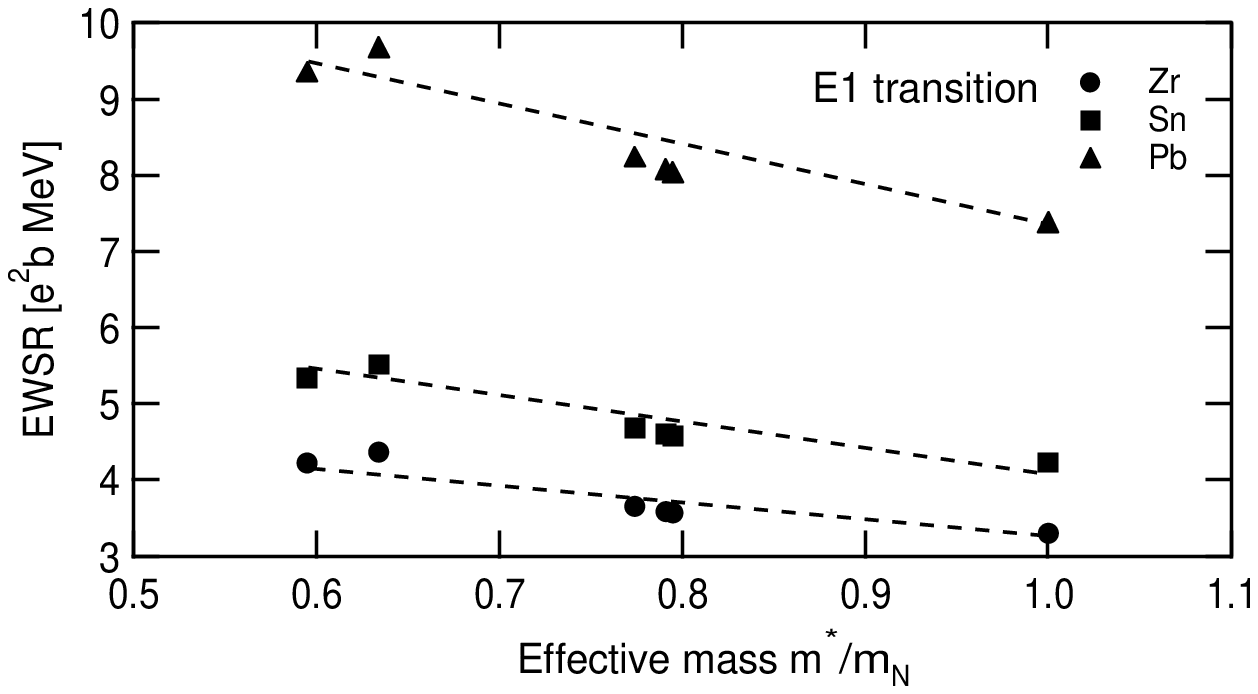}
\includegraphics[width=7.0cm,clip]{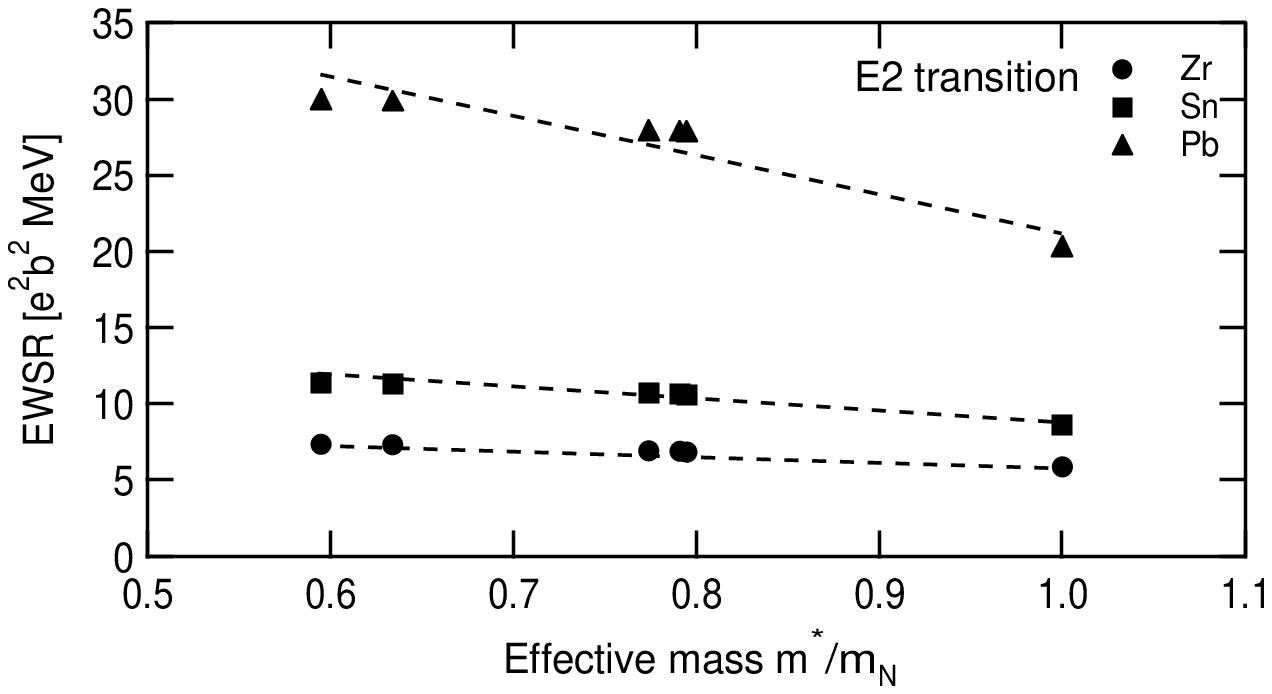}
\caption{\label{EWSRmass}
Energy-weighted sum rules of E0, E1, and E2 transitions
in $^{90}$Zr, $^{116}$Sn, and $^{208}$Pb as a function of the
effective mass $m^*/m_N$ obtained with the parameter sets RHAT1, RHAT2, RHAT3, TM1, and NL3.
The classical values are also plotted in $m^*/m_N=1$.}
\end{figure}

Energy-weighted sum rules depend on the completeness of the particle-hole basis and
are in some case preserved in the RPA. 
Therefore, the EWSR is often used as an indicator of the strength of the nuclear response
in the calculation which requires the summation over all nuclear intermediate states.
The EWSR value for the Coulomb response is given by
\begin{eqnarray}
\sum_{I'} B(E\lambda; 0\rightarrow I') \omega_{I'} \stackrel{q \rightarrow 0}{=} &
\frac{1}{\pi} \frac{2\lambda+1}{(4\pi)^2} \left(\frac{(2\lambda+1)!!}{q^\lambda}\right) 
\int \omega d\omega {\rm Im} \Pi_{RPA} ({\cal Q}\gamma^0,{\cal Q}\gamma^0; q,q;\omega) 
\hspace{3mm} (\lambda \ne 0), \\
\sum_{I'} B(E0; 0\rightarrow I') \omega_{I'} \stackrel{q \rightarrow 0}{=} &
\frac{1}{\pi} \frac{1}{(4\pi)^2} \left(\frac{6}{q^2}\right) 
\int \omega d\omega {\rm Im} \Pi_{RPA} ({\cal Q}\gamma^0,{\cal Q}\gamma^0; q,q;\omega)
\hspace{3mm} (\lambda = 0),
\end{eqnarray}
where $\lambda$ denotes the multipolarity. In the present calculation we take a small momentum transfer $q=10$ MeV.
The results for $\lambda=0$, $1$, and $2$ of $^{90}$Zr, $^{116}$Sn, and $^{208}$Pb 
are shown in Table III, where we take the summation over the RPA states up to 70 MeV for the monopole states 
and 50 MeV for the other states.
From this table, we find that all relativistic results of the EWSR value are somewhat larger than the classical EWSR values of Ref.~\cite{BM75} 
in any multipole states. The reason has been previously presumed as being due to the effective mass\cite{MC89,PR85}.
Certainly, the EWSR values calculated from the RRPA states are approximately proportional to 
the effective mass as depicted in Fig.~\ref{EWSRmass}.
The model including the vacuum polarization cannot shift the effective mass to less than $m^*/m_N = 0.75$ due to the feedback effect
from the vacuum. As a result, the EWSR's are to be suppressed in comparison with the value calculated with TM1 and NL3.
Nevertheless, they are still enhanced by $10-30$\% over the nonrelativistic classical results.
In fact, the excess strength has been reported in the analyses of the giant monopole resonance, where
the single folding (deformed potential) model led to the strength with $154-160$ ($112-129$)\% of the EWSR
in the angular distribution of the differential cross section for the 13.7 keV peak in $^{208}$Pb\cite{YO97}.
The corresponding EWSR for the monopole states in $^{208}$Pb with RHAT1 ($134$\%)
is close to the analysis with the deformed potential model, while that with NL3 ($147$\%)
is close to the analysis with the single folding model.

Although the present results in Table III are thus consistent with the scattering analysis,
this model dependence of the results suggests that the validity of the nuclear model with vacuum polarization has to 
be verified by other experimental data.
In Ref.~\cite{NI05}, $\beta$-decay rates of $r$-process in neutron rich nuclei 
were systematically investigated and it was found that, in order to explain observed data, 
both of the relativistic effective mass, called the Dirac mass there, 
and the empirical effective mass derived from the non-relativistic analysis, 
which is related to the time like component of
the vector interaction, should be considerably larger than those of the typical
relativistic models.
With an additional isoscalar tensor-coupling interaction,
the relativistic effective mass $m^*/m=0.67$, which is considerably higher than that of the typical relativistic models, 
gave a significant improvement in the $\beta$-decay analysis.
Although the experiment demands a larger value, however, 
the effective mass could not be increased more, provided that the ground-state properties of
finite nuclei, including the spin-orbit splittings, are reproduced at the same time.
The present nuclear model, in contrast, generates the relativistic effective mass larger than $m^*/m=0.75$ by means of 
the vacuum-polarization effect.
Inclusion of the vacuum polarization, hence, may resolve these difficulties and support the validity of the present model.
In addition to the $\beta$-decay analysis, it would be interesting to perform the analysis of anomaly in the nuclear-polarization corrections 
for the $^{90}$Zr, $^{208}$Pb, and Sn isotope by the present extended relativistic nuclear models;
the nuclear-polarization correction is considerably sensitive to the EWSR strength and the anomaly seems to require the effective 
mass effect\cite{HA02,HA041}.
The vacuum effect will be verified more through such a systematic study of the effective mass in the relativistic model.


\subsection{Giant monopole resonances}

\begin{figure}[h]
\centering
\includegraphics[width=7.0cm,clip]{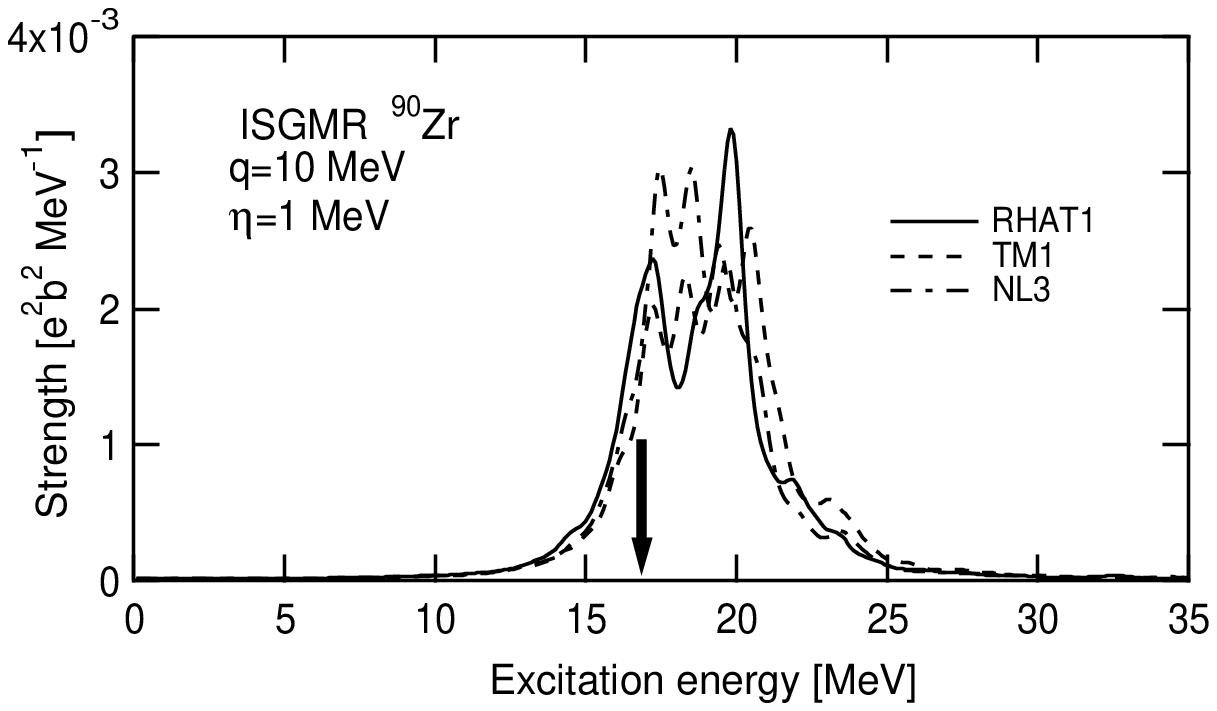}
\includegraphics[width=7.0cm,clip]{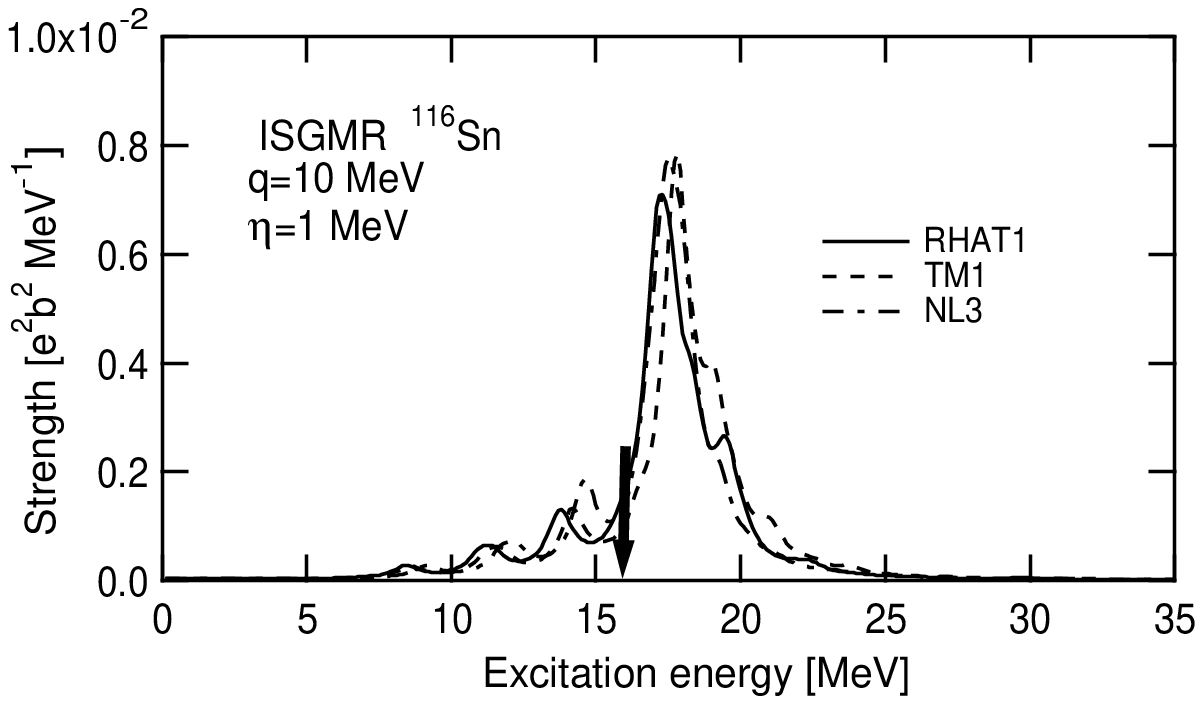}
\includegraphics[width=7.0cm,clip]{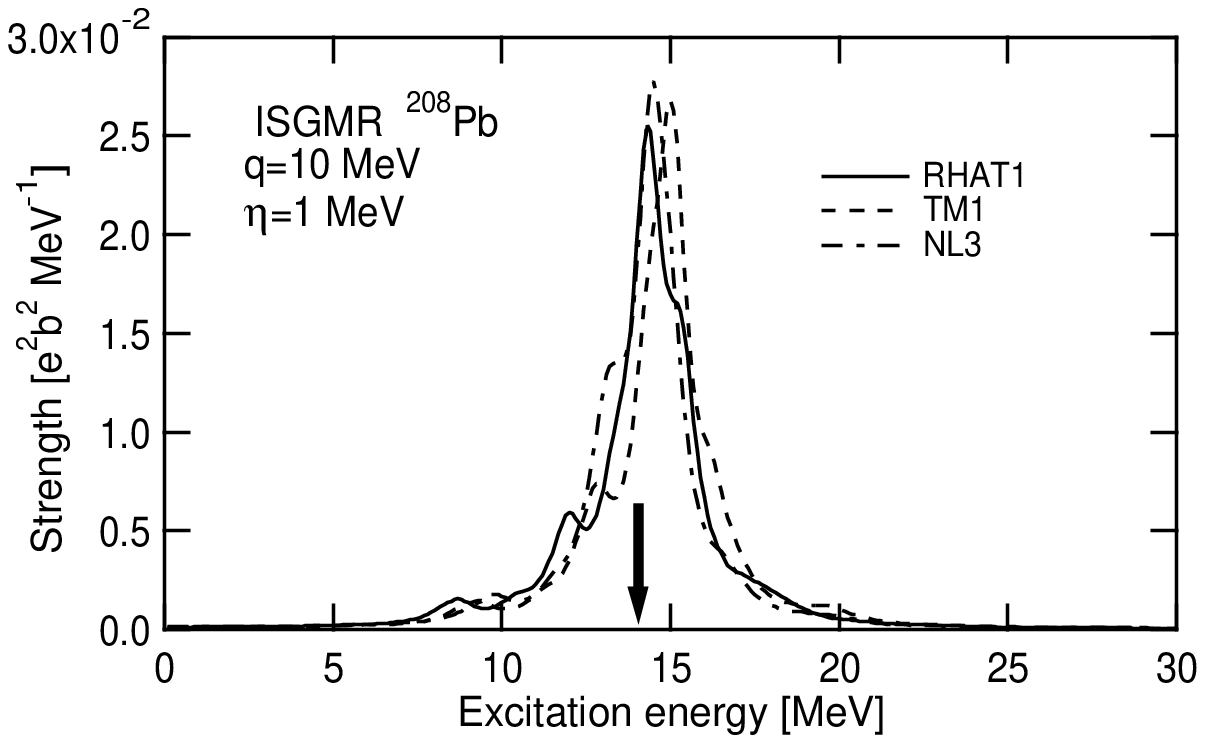}
\caption{\label{ISGMR}
Calculations of the isoscalar-monopole strength in 
$^{90}$Zr, $^{116}$Sn, and $^{208}$Pb.
Arrows indicate experimental peak energies
\cite{YO99,YO041}.}
\end{figure}

\begin{figure}[h]
\centering
\includegraphics[width=7.0cm,clip]{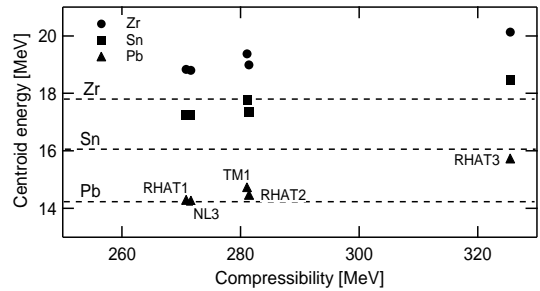}
\caption{\label{ISGMRcom}
The centroid energy of the breathing mode in $^{90}$Zr, $^{116}$Sn, and $^{208}$Pb as a function of the
compression modulus of nuclear matter obtained with the parameter sets RHAT1, RHAT2, RHAT3, TM1, and NL3.
The corresponding data are displayed by the dashed lines\cite{YO99,YO041}.}
\end{figure}

The compressibility of nuclear matter is important in the description of the properties of nuclei.
It is generally accepted that the best procedure to determine the compression modulus
is to calculate isoscalar monopole energies regarded as the "breathing mode" of the nucleus by using the microscopic models.
In the present analysis we examine the predictions of the various parameter sets of the effective Lagrangian 
listed in Table I within the framework of the self-consistent RRPA.
The isoscalar-giant monopole-resonance (ISGMR) 
strength distributions for $^{90}$Zr, $^{116}$Sn, and $^{208}$Pb are displayed in Fig.~\ref{ISGMR} where 
we set the momentum transfer $q=10$ MeV and the width $\eta=1$ MeV. 
The solid curve is the result with the parameter set RHAT1 and the dashed (the dashed-dotted)
one is the result with the parameter set TM1 (NL3). 
The experimental data of excitation energies is also indicated by the arrow\cite{YO99,YO041}.
It is found that the difference in these figures among the parameter sets is small.
This means that the inclusion of the vacuum polarization has a negligible effect in the position of the monopole excitation energy 
if the compressibility is fixed to a certain value.
In Fig.~\ref{ISGMRcom}, we show the ISGMR centroid for these nuclei as a function of the compressibility.
Evidently we see that the centroid energies with the similar compressibility coincide with each other, and are
irrelevant to the inclusion of the vacuum polarization.

Youngblood {\it et al.} obtained the compressibility $K=220-240$ MeV from the analysis of experiment with the various 
non-relativistic calculations\cite{YO99}, while the analysis with ordinary relativistic models gave $K=250-270$ MeV\cite{MA01}.
The origin of the difference between the relativistic and non-relativistic predictions is not understood so far.
The present analysis including the vacuum polarization did not resolve this discrepancy as far as the results of $^{208}$Pb are concerned.
However, the present results for $^{90}$Zr and $^{116}$Sn indicate that the relativistic model may require the lower value for the
compression modulus. For this problem we need further investigations including the pairing force.

\begin{figure}[h]
\centering
\includegraphics[width=7.0cm,clip]{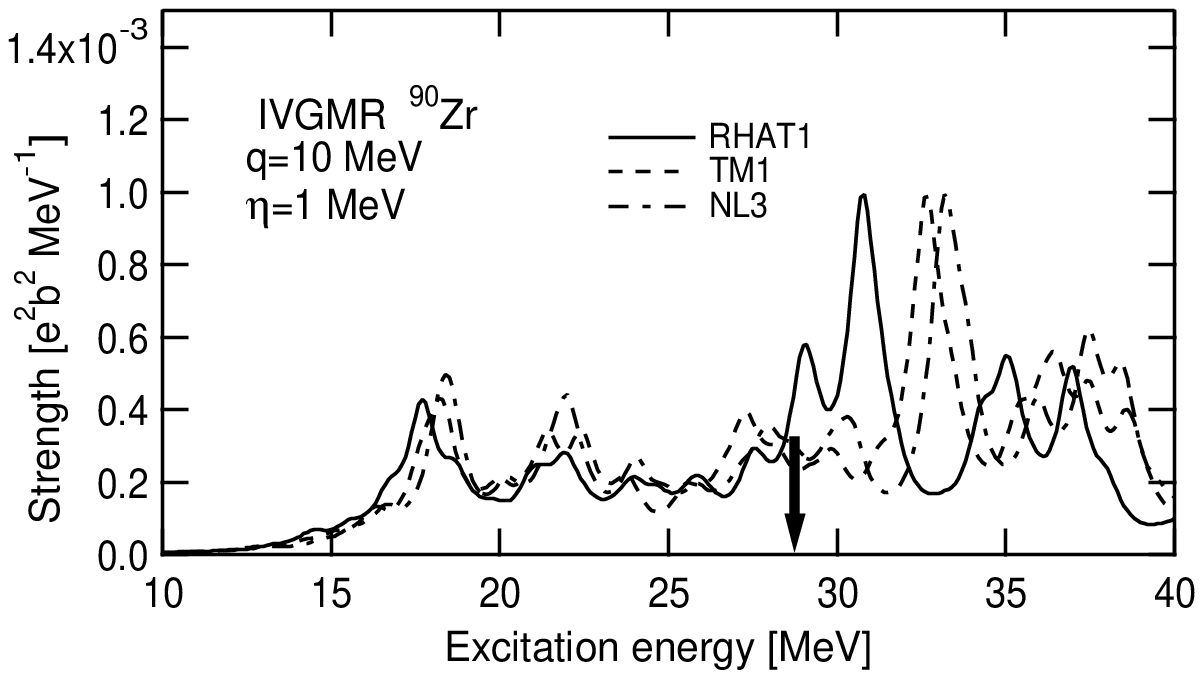}
\includegraphics[width=7.0cm,clip]{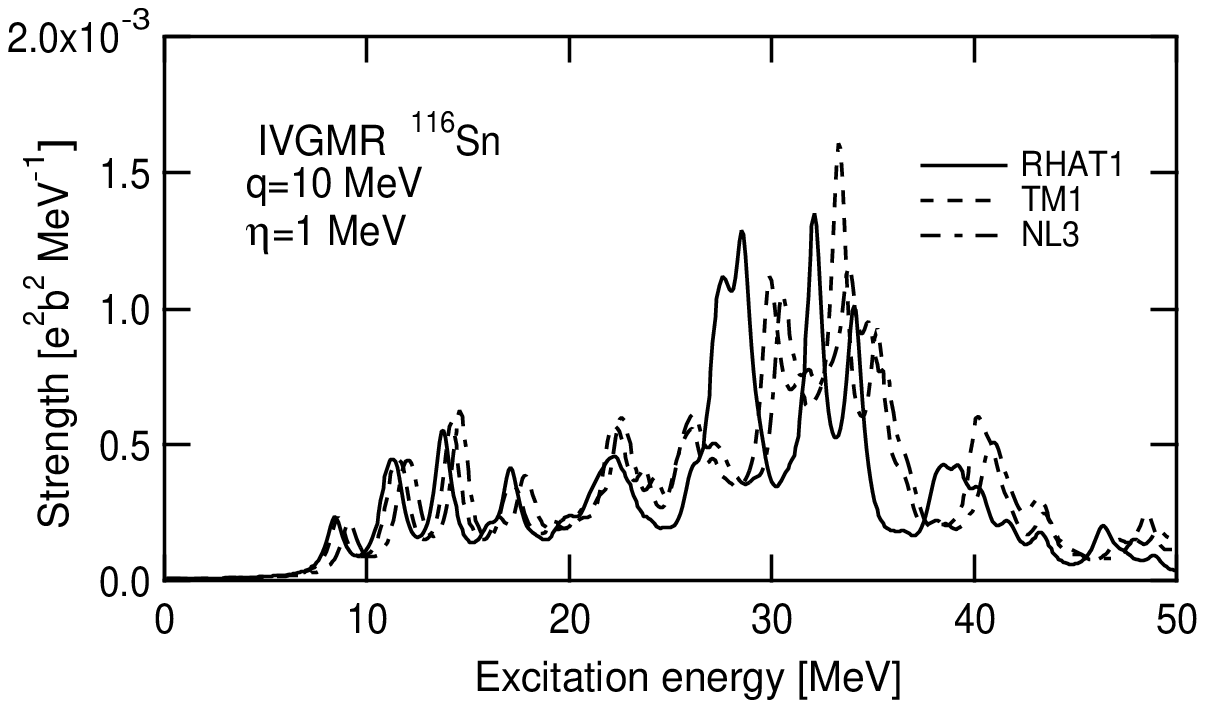}
\includegraphics[width=7.0cm,clip]{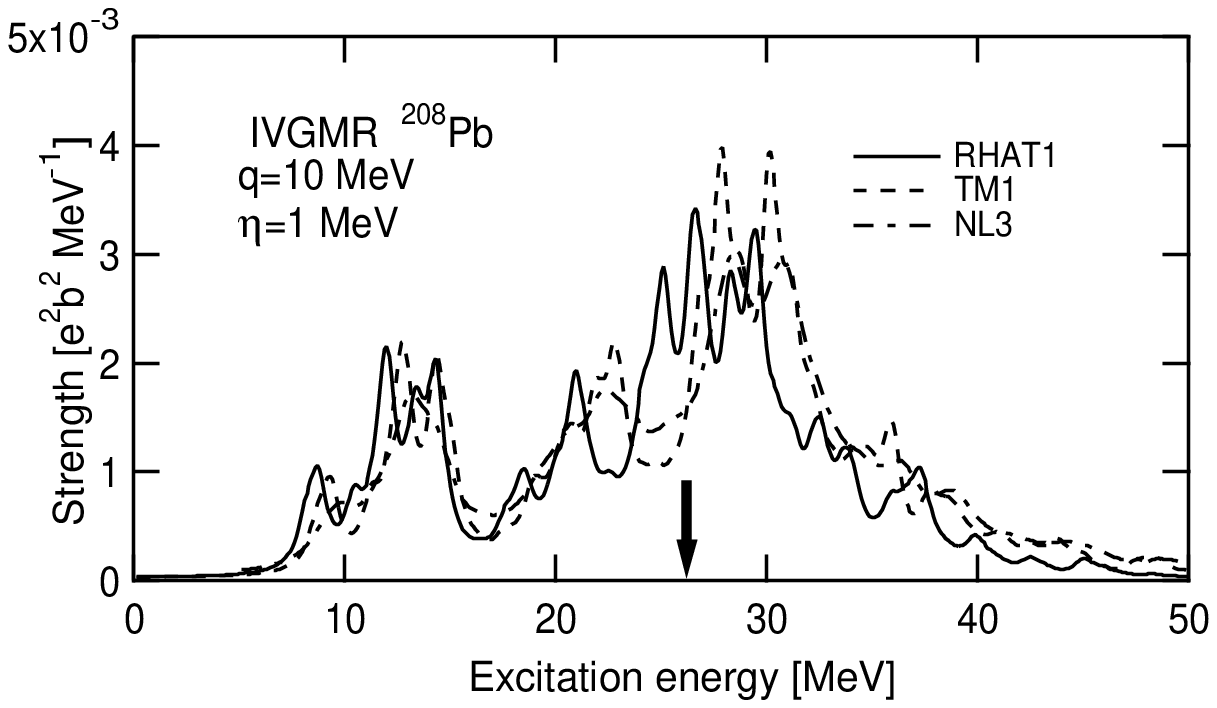}
\caption{\label{IVGMR}
Calculations of the isovector-monopole strength in 
$^{90}$Zr, $^{116}$Sn, and $^{208}$Pb.
Arrows indicate experimental energies
\cite{ER84,BO85}.}
\end{figure}

In contrast to the ISGMR, the studies to locate the isovector giant monopole resonance (IVGMR) have led to ambiguous results
because it is in general found at higher excitation energy with less collectivity.
The calculated IVGMR strengths induced by mainly the $\rho$-meson interaction in the present models are also distributed widely in fragments.
We display the IVGMR strength distributions for $^{90}$Zr, $^{116}$Sn, and $^{208}$Pb in Fig.~\ref{IVGMR} with the three parameter
sets RHAT1, TM1, and NL3. The results do not depend much on the particular model used, and 
the positions of the IVGMR strength are in good agreement with the experiments indicated by the arrow\cite{ER84,BO85}.

\subsection{Giant dipole resonances}

\begin{figure}[h]
\centering
\includegraphics[width=7.0cm,clip]{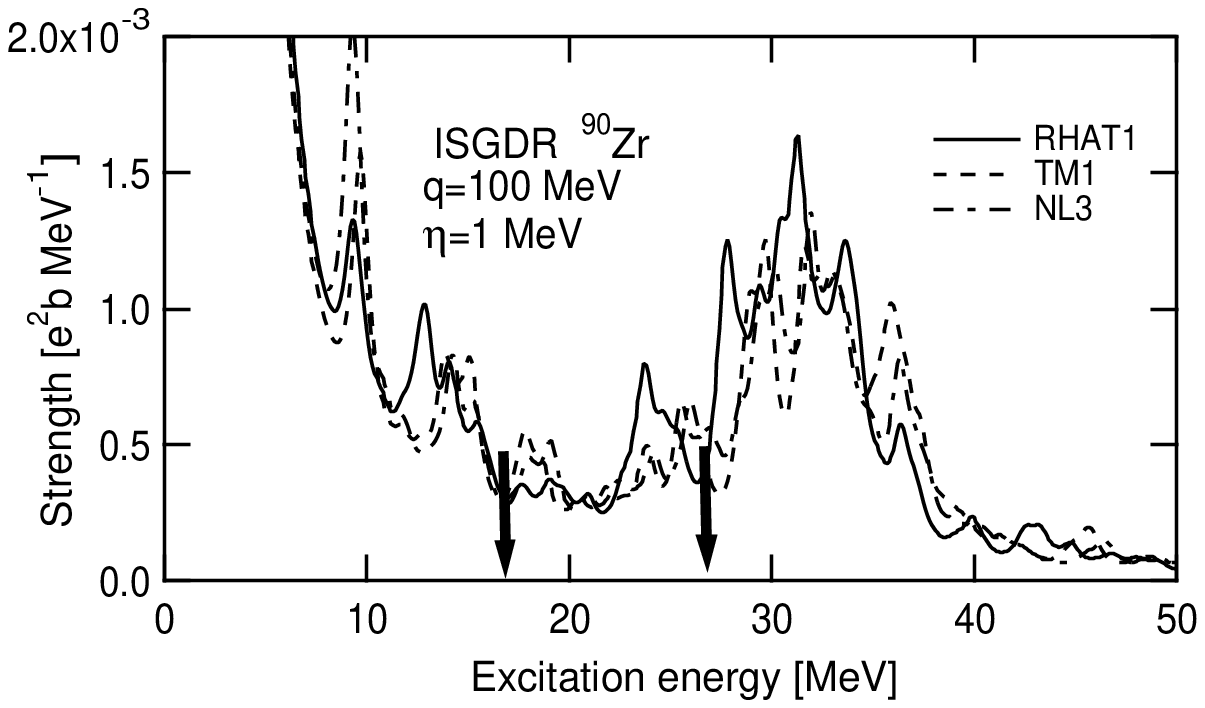}
\includegraphics[width=7.0cm,clip]{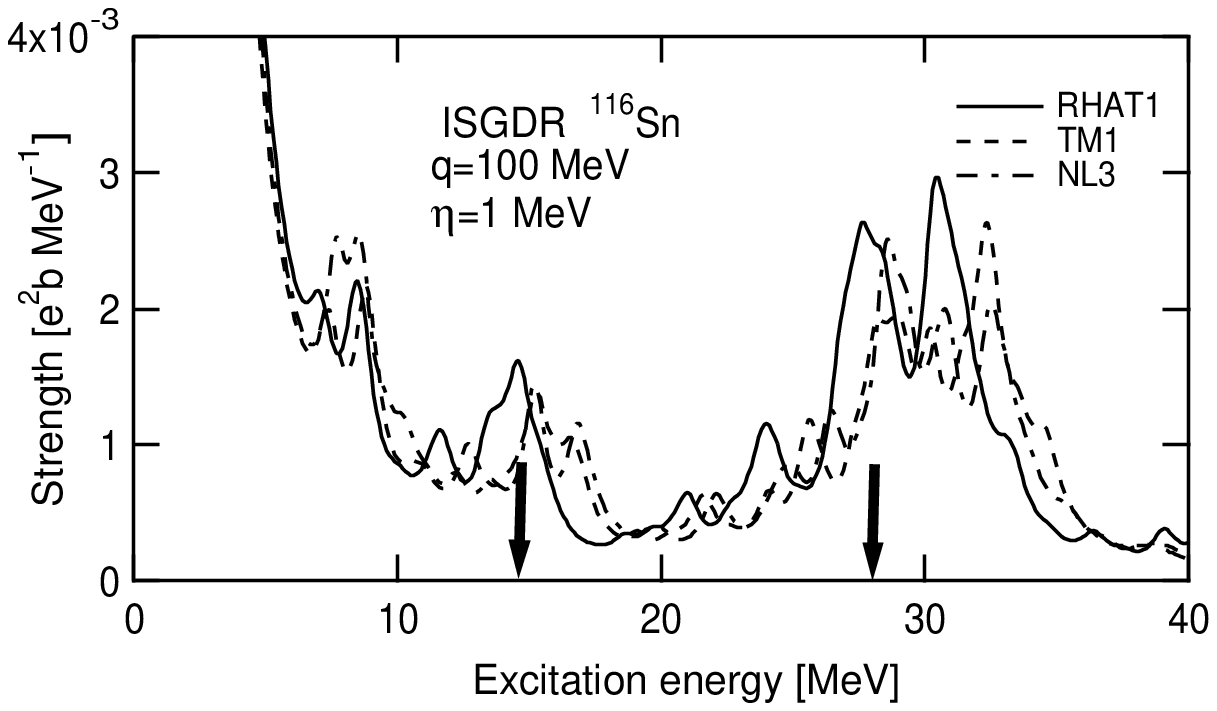}
\includegraphics[width=7.0cm,clip]{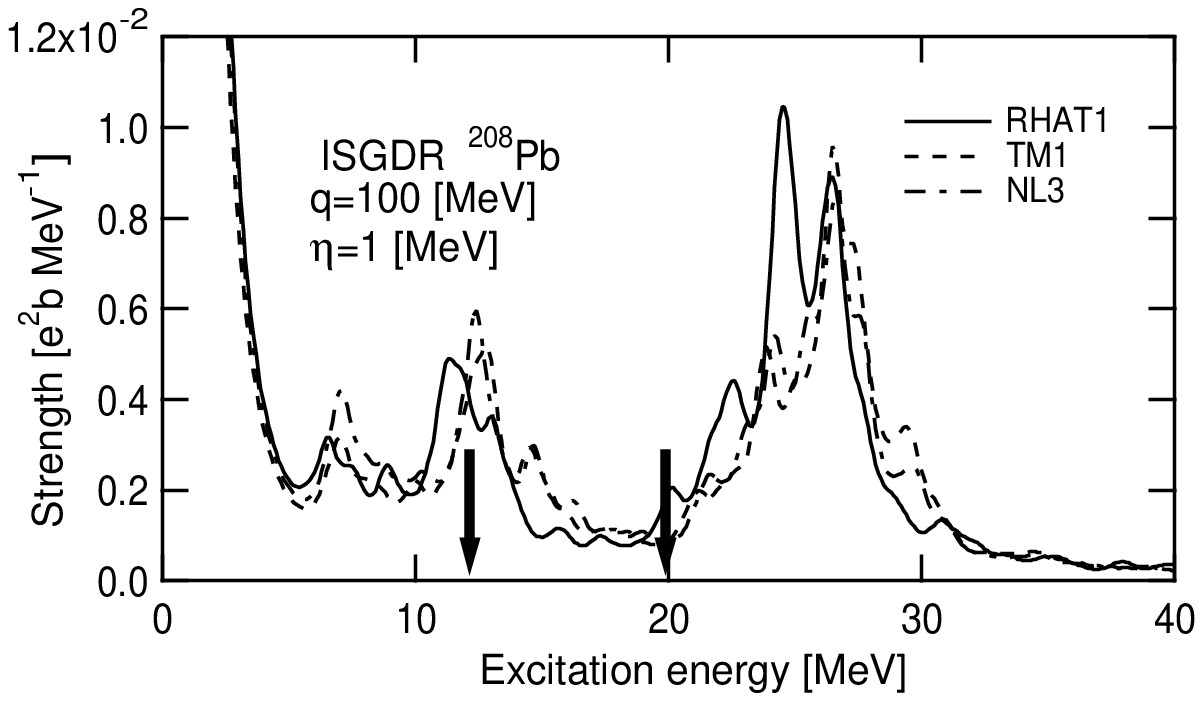}
\caption{\label{ISGDR}
Calculations of the isoscalar-dipole strength in 
$^{90}$Zr, $^{116}$Sn, and $^{208}$Pb.
Arrows indicate experimental energies
\cite{CL01,YO041}.}
\end{figure}

The isoscalar-giant dipole resonance (ISGDR) contains the spurious component caused by the center-of-mass motion.
In the previous section, we indicated that the spurious state appears at zero excitation energy 
with the fully self-consistent calculation, but the physical excitations with $q=10$ MeV become negligibly small
since the spurious state has an exceedingly strong collectivity at low-momentum transfer.
In order to identify the physical excitations, hence, we employ a larger momentum transfer, $q=100$ MeV, in the ISGDR mode.
The ISGDR strengths for $^{90}$Zr, $^{116}$Sn, and $^{208}$Pb are depicted in Fig.~\ref{ISGDR}.
Experimentally, it is observed to split into upper- and lower-energy components in the ISGDR strengths
(the experimental peaks are indicated by two arrows in the figure\cite{CL01,YO041}).
Both in the model with (the solid curve) and without the vacuum polarization (the dashed and dash-dotted curves), 
we see that there are two components similar to observed ones.
In comparison with the experiments, however, the splittings between the upper and lower peaks are somewhat large in all calculations employed here.
On the other hand, we find that calculations reproduce the peaks of the lower-energy region in $^{116}$Sn and $^{208}$Pb.
Moreover, the results with the vacuum polarization for the upper component are different from other models, and
approach to the data, although the predicted peaks are still $2-5$ MeV higher than the experimental ones.
It has been reported that the upper component is the compression mode, whose energy depends on the compression modulus,
while the energy of the lower component changes with mass but is essentially independent of compression modulus\cite{CO00,VE00}.
However, in the present model with the parameter set RHAT1, which has a similar compression modulus to that with TM1 and NL3 (see Table I),
the energies in upper region for all nuclei certainly shift to the lower energy than those with TM1 and NL3.
This suggests that not only the compression modulus but also
the shell structure is important to resolve the discrepancy in the position of the peak for the upper-energy component.

\begin{figure}[h]
\centering
\includegraphics[width=7.0cm,clip]{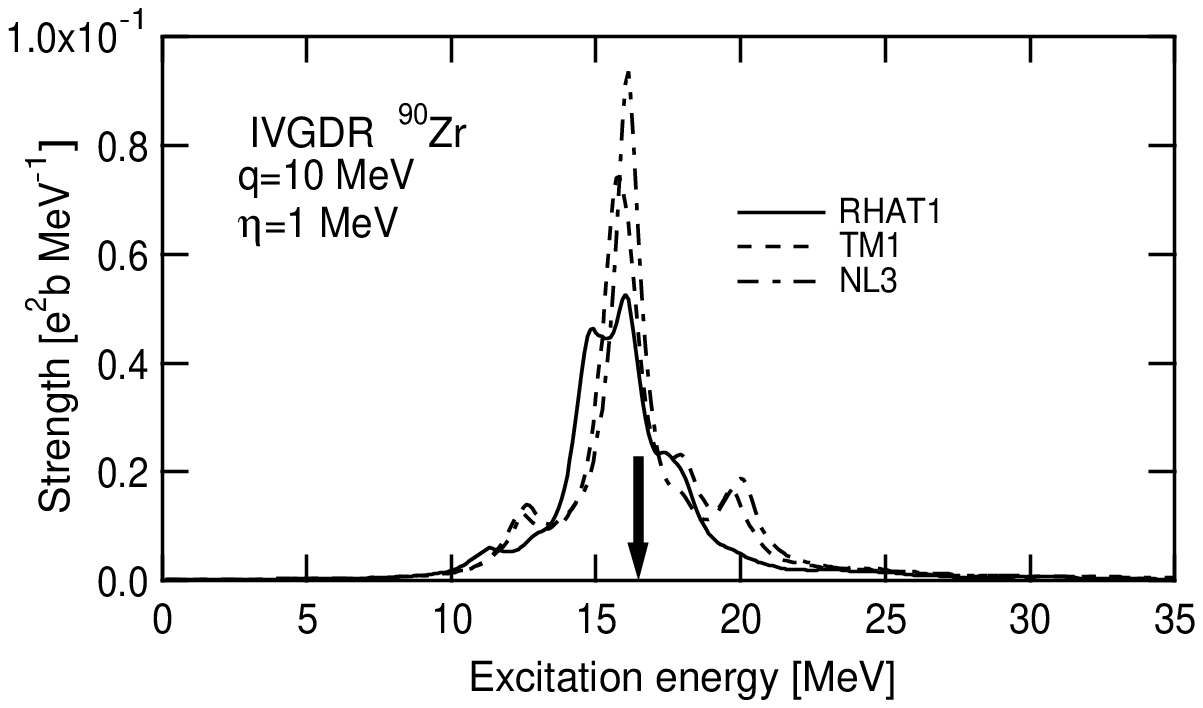}
\includegraphics[width=7.0cm,clip]{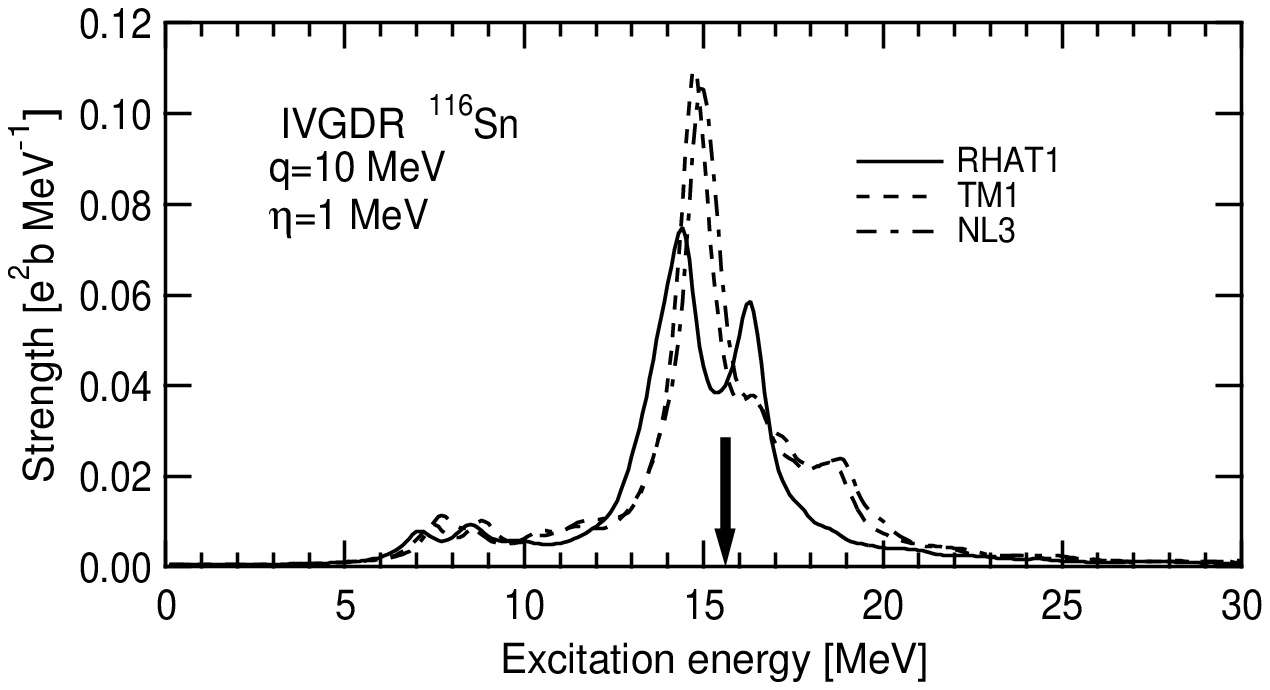}
\includegraphics[width=7.0cm,clip]{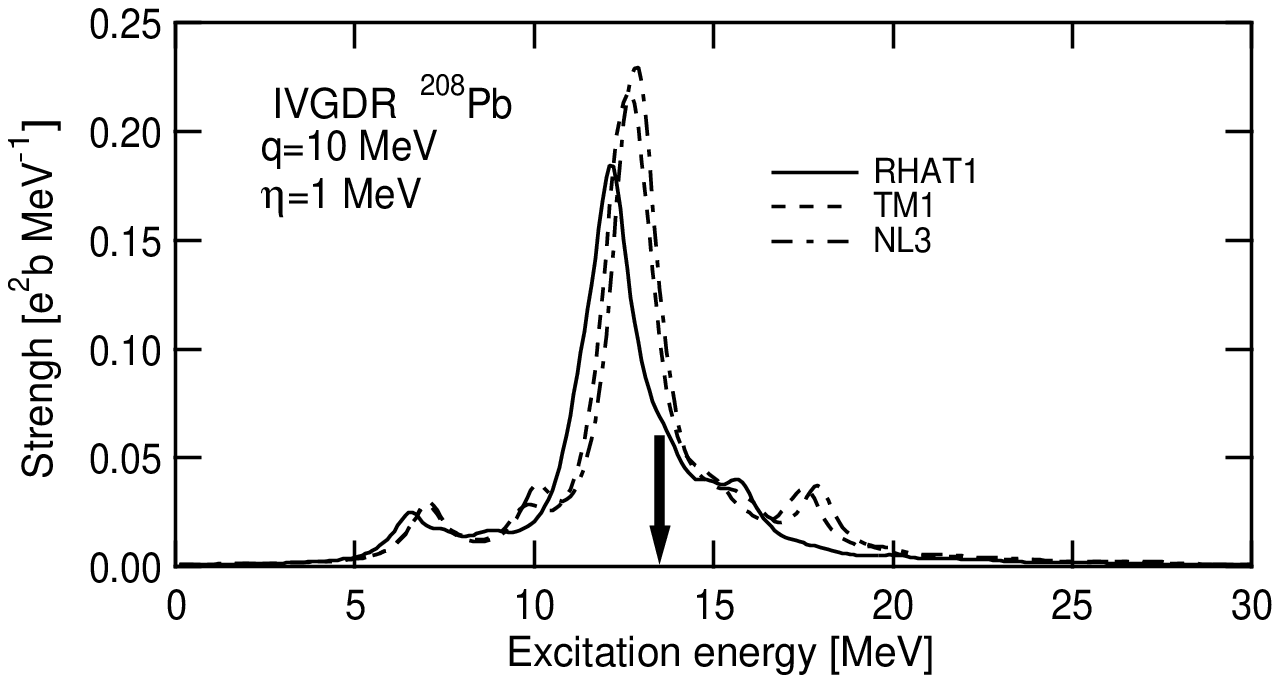}
\caption{\label{IVGDR}
Calculations of isovector-dipole strength in 
$^{90}$Zr, $^{116}$Sn, and $^{208}$Pb.
Arrows indicate experimental energies
\cite{BE75}.}
\end{figure}

The isovector giant dipole resonance (IVGDR) are also illustrated in Fig.~\ref{IVGDR}.
There, we find that due to the effective mass suppressed by the vacuum polarization, the strengths of the RHAT1 (the solid line) become weak for all nuclei.
The centroid energy of $^{90}$Zr with RHAT1 ($16.4$ MeV) is in agreement with the experimental data ($16.5 \pm 0.2$ MeV\cite{BE75})
while the energies of $^{116}$Sn ($15.1$ MeV) and $^{208}$Pb ($12.9$ MeV) are slightly smaller than the experimental values
($15.7 \pm 0.2$ MeV for $^{116}$Sn and $13.5 \pm 0.2$ MeV for $^{208}$Pb\cite{BE75}).
However, the calculated centroid energies for $^{116}$Sn and $^{208}$Pb
contain contributions from the low-energy discrete states (Fig.~\ref{IVGDR}), which were not considered in the analysis of
the experiment. Excluding these contributions,
both of the centroid energies ($15.6$ MeV for $^{116}$Sn and $13.4$ MeV for
$^{208}$Pb) are also in agreement with the experiments.

\subsection{Giant quadrupole resonances}

The giant quadrupole resonance is induced by the $2\hbar\omega$ particle-hole correlation 
in the harmonic oscillator potential model, and the isoscalar giant quadrupole resonance (ISGQR)
generates a collective state in the lower-energy region while the isovector giant quadrupole resonance (IVGQR)
generates it in the higher-energy region.
In relativistic models the shell structure is significantly affected by the effective mass. 
As a result, the models with and without the vacuum polarization may yield
the different results from each other both in the ISGQR and the IVGQR.

\begin{figure}[h]
\centering
\includegraphics[width=7.0cm,clip]{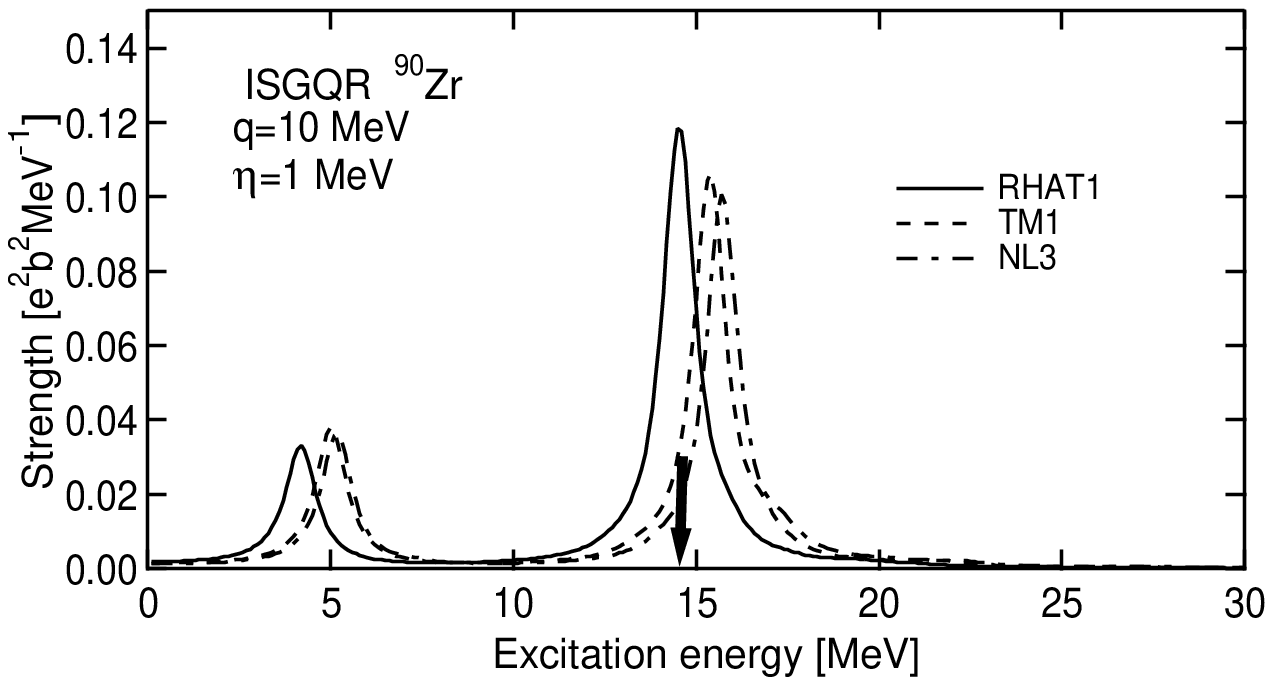}
\includegraphics[width=7.0cm,clip]{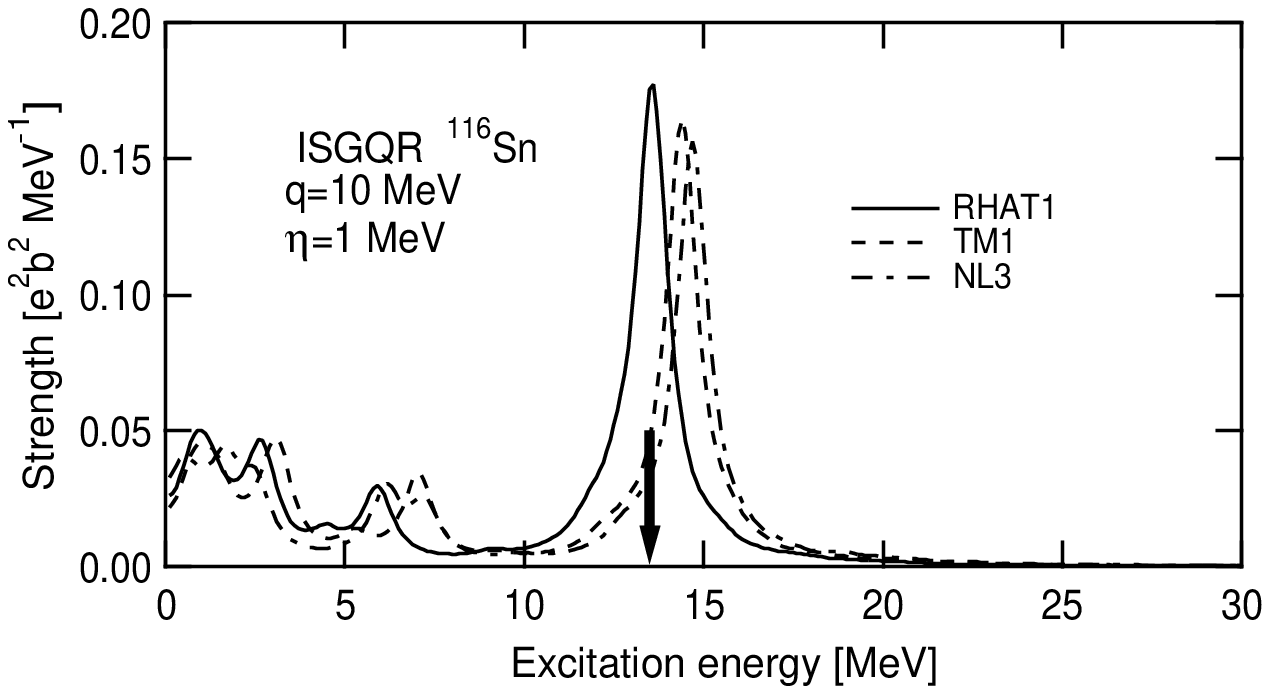}
\includegraphics[width=7.0cm,clip]{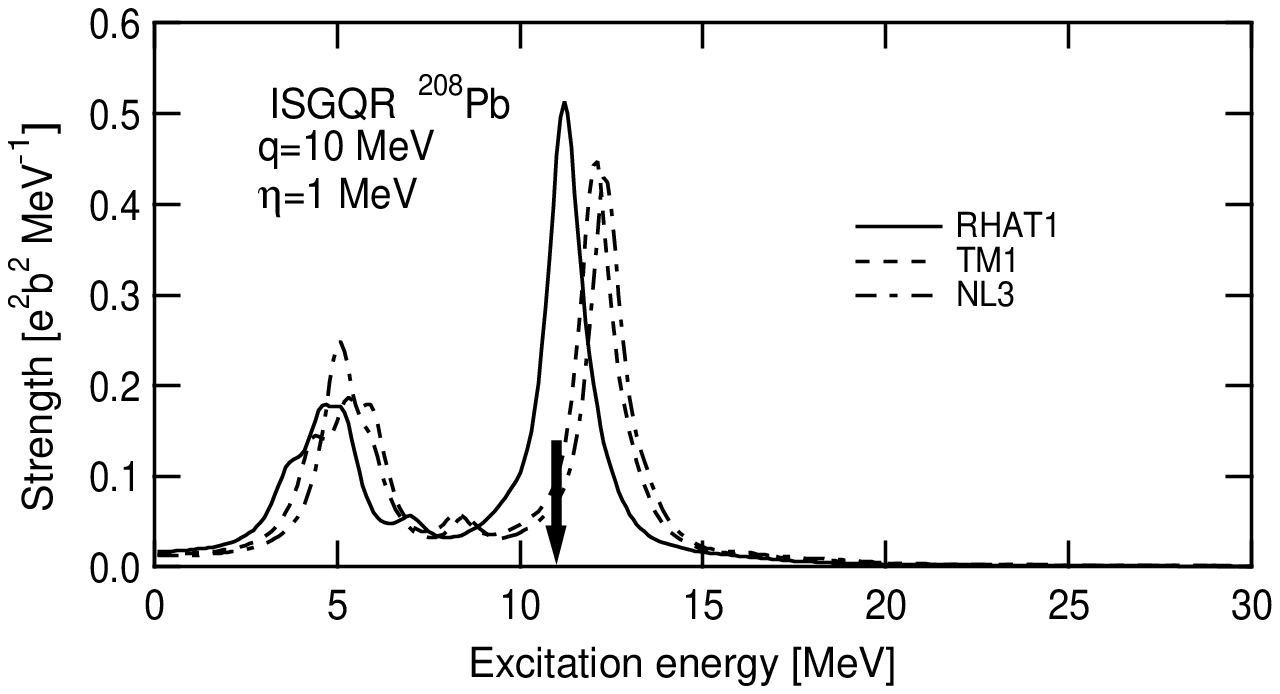}
\caption{\label{ISGQR}
Calculations of isoscalar-quadrupole strength in 
$^{90}$Zr, $^{116}$Sn, and $^{208}$Pb.
Arrows indicate experimental energies
\cite{YO041,YO042}.}
\end{figure}

\begin{figure}[h]
\centering
\includegraphics[width=7.0cm,clip]{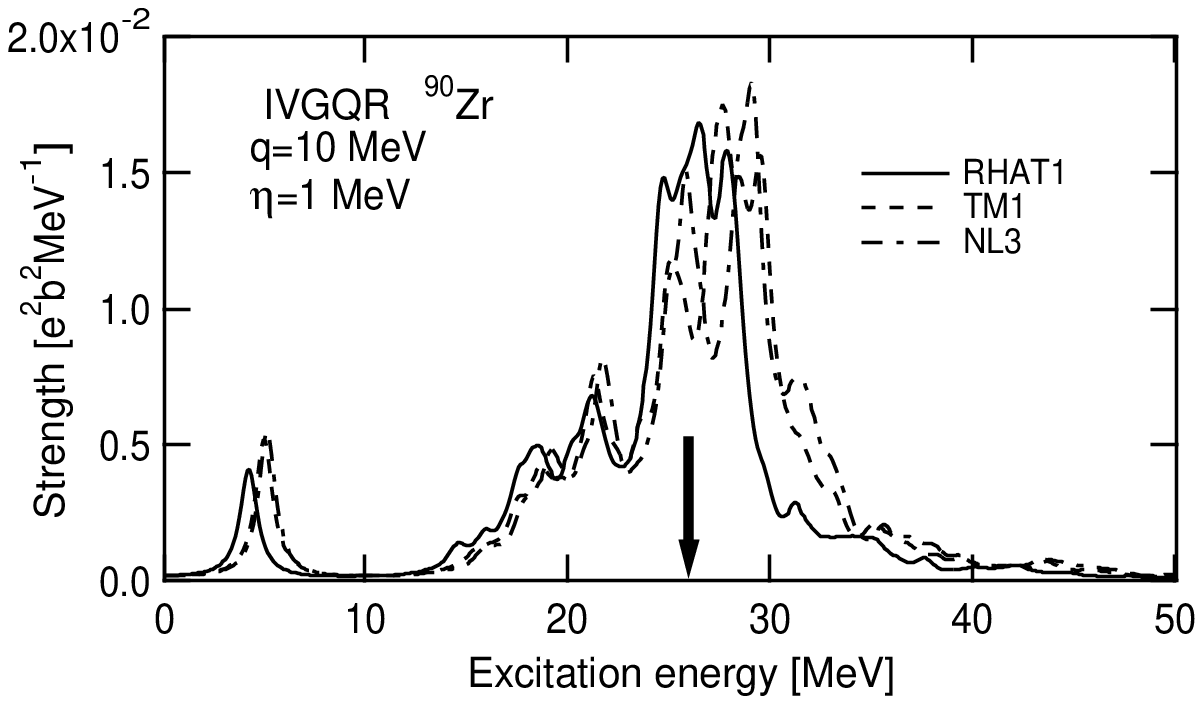}
\includegraphics[width=7.0cm,clip]{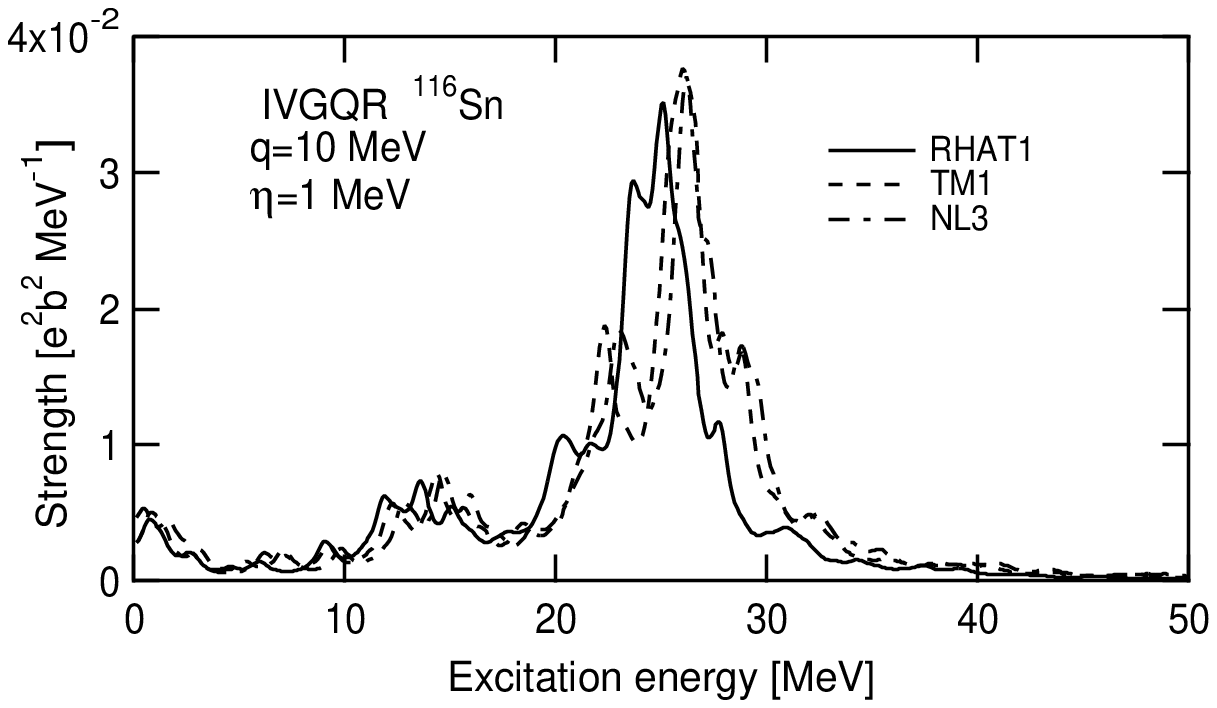}
\includegraphics[width=7.0cm,clip]{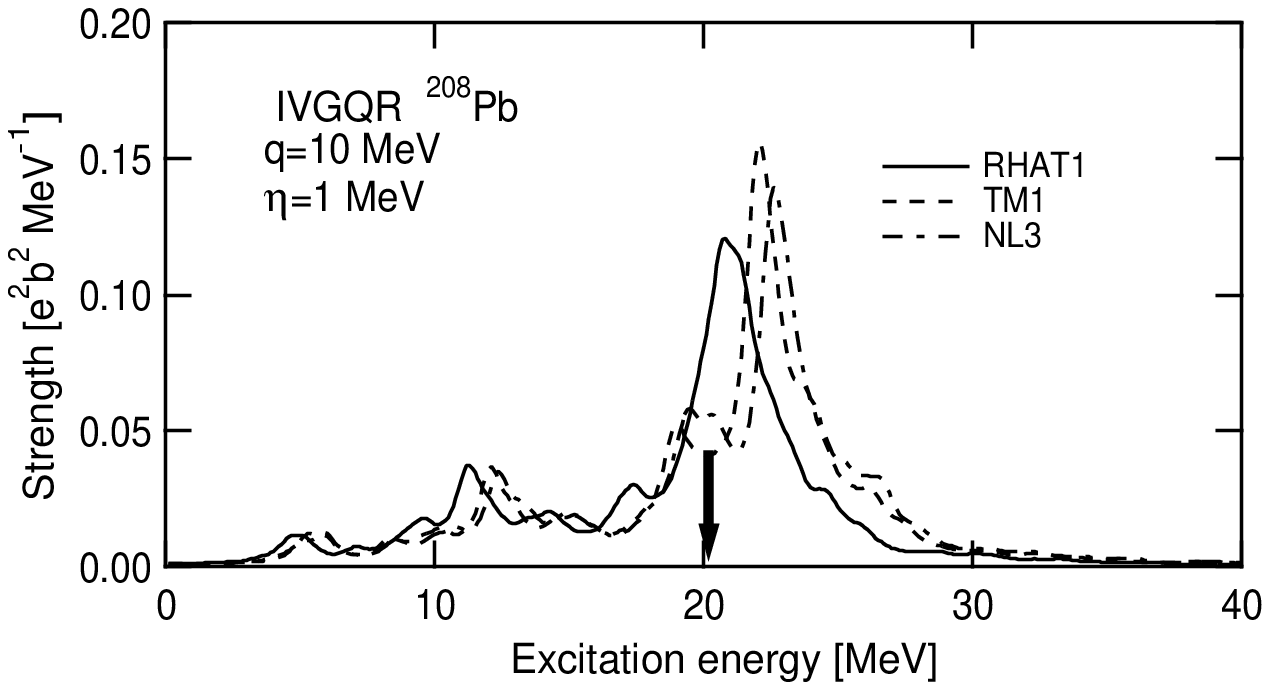}
\caption{\label{IVGQR}
Calculations of isovector-quadrupole strength in 
$^{90}$Zr, $^{116}$Sn, and $^{208}$Pb.
Arrows indicate experimental energies
\cite{DA92,GO94}.}
\end{figure}

\begin{figure}[h]
\centering
\includegraphics[width=7.0cm,clip]{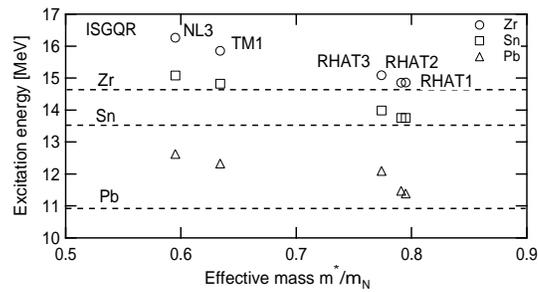}
\caption{\label{GQRmass}
Excitation energy of the isoscalar-quadrupole resonance
of $^{90}$Zr, $^{116}$Sn, and $^{208}$Pb is plotted as a function of the
effective mass $m^*/m_N$ obtained with the parameter sets RHAT1, RHAT2, RHAT3, TM1, and NL3.
The corresponding data are displayed by the dashed lines
\cite{YO041,YO042}.}
\end{figure}

The EWSR distributions of the Coulomb response of the ISGQR (IVGQR) mode are shown in Fig.~\ref{ISGQR} (Fig.~\ref{IVGQR})
for $^{90}$Zr, $^{116}$Sn, and $^{208}$Pb. The ISGQR strength exhibits a strong collectivity.
The peaks seen in the region of less than 8 MeV for the ISGQR mode come from the low-lying discrete states.
In all nuclei, the downward shift of the model with vacuum polarization (RHAT1) from the model without vacuum polarization 
(TM1 and NL3) is clearly observed both in the ISGQR and the IVGQR modes.
Also, it is found that the RHAT1 results are in excellent agreement with the experimental data indicated by the arrow for each nucleus\cite{YO041,YO042,DA92,GO94}.
We show the RHAT1 results for the centroid energies of the ISGQR in Table IV where they are compared with the corresponding data.
The centroid energies calculated by the relativistic model with the parameter set NL3 and the relativistic point-coupling model 
were approximately 1-2 MeV above the experimental energies\cite{YO041,NI052}.  
These disagreements are drastically improved by the present model including the vacuum polarization.
Thus, the effective mass plays an important role in the improvement of the results in the ISGQR, while not in the ISGMR
with compressibility fixed to a certain value.
The centroid energies of the ISGQR as a function of the effective mass are plotted in Fig.~\ref{GQRmass}, where
we see that the effective mass, as expected, has a substantial effect on the energy of the ISGQR.
In the present model the large effective mass is theoretically caused by the vacuum effect.
It is thus shown that the inclusion of the vacuum polarization gives much better results for nuclear excitations through the effective mass effect.

\section{Summary}

In summary, we have studied the nuclear excitations by the self-consistent RRPA method with the vacuum-polarization contribution
extended with the tensor-coupling term for the $\omega$ meson.
We have shown that the vacuum polarization can be easily and duly included in the RPA calculation if the derivative expansion 
for the one-loop nucleon contribution is taken into account in the effective Lagrangian and if the vacuum-polarization parts 
in the unperturbed-polarization tensor are consistent with the second-order fluctuation of its Lagrangian.
We have indicated numerically that the vacuum-polarization contribution is quite important to remove the center-of-mass motion 
from the isoscalar dipole resonance, and the self-consistent calculation involving the Schwinger term
conserves the vector current in the multipole transition.

Using this method, we have investigated the properties of the collective excitations of $^{90}$Zr, $^{116}$Sn, and $^{208}$Pb.
With the present model, it is impossible to obtain the effective mass as small as that of the conventional relativistic model, 
since the vacuum polarization works to reduce the scalar-meson field.
Therefore, the enhancement of the EWSR strength due to the effective mass is suppressed by the inclusion of the vacuum polarization.
This fact does not contradict with the scattering data and is consistent with the analysis of the $\beta$-decay lifetimes. 
In addition, we stress here that the present extended relativistic model is able to give better predictions in almost all giant resonances
than the relativistic model without the vacuum polarization.
In particular, we have shown that the calculated ISGQR centroids reproduce the experimental data for each nucleus nicely.
Thus the present study suggests that the vacuum-polarization effect plays an important role in the excitations through the change of 
the single-particle structure by means of the effective mass effect.
Further calculations testing the nuclear models for observables with respect to nuclear excitations like a nuclear-polarization 
correction in muonic atoms would be very interesting.

Finally, we mention the conflict between the relativistic and non-relativistic models for the compression modulus.
As for the ISGMR in $^{208}$Pb, the estimation that the compressibility in nuclear matter calculated 
by the relativistic model is in the range $K=250-270$ MeV is a likely consequence.
However, for the ISGMR in $^{90}$Zr and $^{116}$Sn, and for the high-energy part of the ISGDR in the present analysis,
the compression modulus seems to be a lower value, and therefore the relativistic result may coincide with the non-relativistic one.
Unfortunately, we could not find a stable solution less than $K=270$ MeV with the vacuum polarization.
In order to resolve this problem, we need the $(\omega_\mu\omega^\mu)^2$ term originated by the parameter set TM1.
The energy of the ISGMR compression modes is not sensitive to the inclusion of the vacuum polarization, that is, to the value 
of the effective mass, which is also the conclusion of the non-relativistic study\cite{BL95}.
On the other hand, the energy peaks of the upper component in the ISGDR, identified as the compression mode, 
shift to the lower energy and approach to the corresponding data by the inclusion of the vacuum polarization.
Therefore, we should study further the relation between the compressibility and the effective mass.

\appendix
\section{Current conservation with tensor-coupling interaction}

In order to prove the current conservation for the RRPA transition densities
in the present model including the tensor-coupling interaction between the $\omega$ meson and nucleons,
we introduce the simplified Lagrangian density,
\begin{align}
\label{Lappen}
\mathcal{L}=&\bar{\psi} 
\left[i\gamma_\mu
\partial^\mu-m_N-g_\omega\gamma^\mu\omega_\mu
-\frac{f_\omega}{4m_N}\sigma^{\mu\nu}(\partial_{\mu}\omega_{\nu}-\partial_{\nu}\omega_{\mu})
-e{\cal Q}\gamma^\mu A_\mu
\right]
\psi
\nonumber\\
&
-\frac{1}{4}(\partial_{\mu}\omega_{\nu}-\partial_{\nu}\omega_{\mu})^2
+\frac{1}{2}m_\omega^2(\omega_{\mu})^2
-\frac{1}{2}(\partial^\mu \omega_\mu)^2
\nonumber\\
&
-\frac{1}{4}(\partial_{\mu} A_{\nu}-\partial_{\nu} A_{\mu})^2
-\frac{1}{2}(\partial^\mu A_\mu)^2,
\end{align}
where the gauge fixing terms, $-1/2(\partial^\mu \omega_\mu)^2$ and $-1/2(\partial^\mu A_\mu)^2$,
are explicitly included.
The Noether's currents of the photon and $\omega$ meson fields, $j_A^\mu$ and $j_\omega^\mu$, 
are same as Eqs.~(\ref{JA}) and (\ref{JO}), respectively.
Using this Lagrangian, the canonical conjugate of the $\omega$-meson field reads as,
\begin{align}
\label{piappen}
\pi^\mu_\omega=\frac{\partial {\cal L}}{\partial \dot{\omega}_\mu}
=(\partial^\mu \omega^0-\partial^0 \omega^\mu)
-\frac{f_\omega}{2m_N}\bar{\psi} \sigma^{0\mu}\psi
-(\partial^\nu \omega_\nu)\delta^{\mu 0},
\end{align}
which depends on the nucleon field through the tensor-coupling term.

Now, we shall define the propagator connecting between the photon and the $\omega$ meson,
$D^{\mu\nu}_{A\omega}(x-y)$;
\begin{align}
\label{DOAppen1}
D^{\mu\nu}_{A\omega}(x-y)
=
\int dx'dy' 
D^{\mu\xi}_{A 0}(x-x')
\Pi_D ({\cal Q}\gamma_\xi,\gamma_\zeta+\frac{f_\omega}{2m_N}\overleftrightarrow{\partial}^\lambda \sigma_{\zeta\lambda};x',y')
D^{\zeta\nu}_{\omega 0}(y'-y),
\end{align}
where $D^{\mu\nu}_{A 0}(x-y)$ and $D^{\mu\nu}_{\omega 0}(x-y)$ are the free propagator of the photon
and $\omega$ meson, respectively.
Then, the four-divergence of the photon-$\omega$ meson propagator gives 
\begin{align}
\label{DOAppen2}
\square^x iD^{\mu\nu}_{A\omega}(x-y)
=
\square^x 
\langle 0| T[A^\mu(x) \omega^\nu(y)]|0 \rangle
=
\langle 0| T[j_A^\mu(x) \omega^\nu(y)]|0 \rangle,
\end{align}
where we use the Euler-Lagrange equation $\square A^\mu(x) = j_A^\mu (x)$.
Substituting Eq.~(\ref{DOAppen1}) into Eq.~(\ref{DOAppen2}) and multiplying $\square^y+m_\omega^2$, 
we obtain the following relation;
\begin{align}
\label{PiHNA}
i\Pi_D ({\cal Q}\gamma^\mu,\gamma^\nu+\frac{f_\omega}{2m_N}\overleftrightarrow{\partial}_\xi\sigma^{\nu\xi};x,y)
=
\langle 0| T[j_A^\mu(x) j_\omega^\nu(y)]|0 \rangle
-\langle 0| [j_A^\mu(x), \dot{\omega}^\nu(y)]_0|0 \rangle,
\end{align}
where we also use the equation $(\square+m_\omega^2) \omega^\mu(x) = j_\omega^\mu (x)$.
From Eq.~(\ref{piappen}) and the quantization condition imposing $[j_A^\mu(x), \pi_\omega^\nu(y)]_0=0$, 
we can easily deduce the relation
\begin{align}
\label{PiHN2}
i\Pi_D ({\cal Q}\gamma^\mu,\gamma^\nu+\frac{f_\omega}{2m_N}\overleftrightarrow{\partial}_\xi\sigma^{\nu\xi};x,y)
=
\langle 0| T[j_A^\mu(x) j_\omega^\nu(y)]|0 \rangle
+\frac{f_\omega}{2m_N}\langle 0| [j_A^\mu(x), \bar{\psi}(y)\sigma^{0\nu}\psi(y)]_0|0 \rangle,
\end{align}
which is exactly same as Eq.~(\ref{PiHN}).
Thus, the straightforward treatment of the Lagrangian including the tensor-coupling term
naturally introduce the Schwinger term, which is the second term in the right-hand side of above equation\cite{AD68}.
The Schwinger term plays a crucial role for the current conservation.
Multiplying Eq.~(\ref{PiHN2}) by $\partial_\nu ^y$, one obtains
\begin{align}
\label{commutator}
\partial_{\nu}^y
i\Pi_D ({\cal Q}\gamma^\mu,\gamma^\nu+\frac{f_\omega}{2m_N}\overleftrightarrow{\partial}_\xi\sigma^{\nu\xi};x,y)
=
-\langle 0| [j_A^\mu(x), j_\omega^0(y)]_0|0 \rangle
+ \frac{f_\omega}{2m_N}
\partial_{\nu}^y\langle 0| [j_A^\mu(x), \bar{\psi}(y)\sigma^{0\nu}\psi(y)]_0|0 \rangle.
\end{align}
The current operators $j_\omega^\mu$ do not commute each other due to the anomalous part, 
and therefore this contribution results in the violation of the conservation law
when the conventional polarization tensor as indicated in Eq.~(\ref{PiH}) only is considered.
It can be easily seen that the Schwinger term cancels the non-vanishing term of the 
$[j_A^\mu(x), \rho_\omega(y)]$ showing the current conservation of the polarization tensor.

\clearpage \setlength{\baselineskip}{0.0in}

TABLE I.  Parameter sets of the effective Lagrangians
\vspace{1mm}\\
\begin{math}
\begin{array}{lccccc}
\hline\hline &
\hspace{7mm}{\rm RHAT1} \hspace{7mm} & 
\hspace{7mm}{\rm RHAT2} \hspace{7mm} & 
\hspace{7mm}{\rm RHAT3} \hspace{7mm} & 
\hspace{7mm}{\rm TM1}   \hspace{7mm} & 
\hspace{7mm}{\rm NL3}   \hspace{7mm} \\
\hline
m_N {\rm [MeV]}
& 938.0
& 938.0
& 938.0
& 938.0
& 939.0\\
m_\sigma {\rm [MeV]}
& 390.0
& 400.0
& 450.0
& 511.198
& 508.194\\
m_\omega {\rm [MeV]}
& 783.0
& 783.0
& 811.0
& 783.0
& 782.501\\
m_\rho {\rm [MeV]}
& 763.0
& 763.0
& 763.0
& 763.0
& 763.0\\
g_\sigma
& 6.046
& 6.226
& 7.085
& 10.0289
& 10.217\\
g_\omega
& 8.264
& 8.407
& 9.220
& 12.6139
& 12.868\\
g_\rho
& 10.285
& 10.055
& 9.155
& 9.2644
& 8.948\\
g_2 {\rm [fm^{-1}]}
& 19.75
& 19.39
& 18.54
& 14.465
& 20.862\\
g_3 
& -19.42
& -19.42
& -20.82
& 3.7098
& -173.31\\
c_3 
& 0.0
& 0.0
& 0.0
& 71.3075
& 0.0\\
\kappa_\omega/m_N 
& 2.9
& 2.8
& 2.5
& 0.0
& 0.0\\
k_F {\rm [fm^{-1}]}
& 1.3021
& 1.2993
& 1.2894
& 1.2905
& 1.2995\\
E/A {\rm [MeV]}
& -16.522
& -16.487
& -16.204
& -16.265
& -16.243\\
m^*/m_N 
& 0.795
& 0.791
& 0.774
& 0.634
& 0.595\\
K {\rm [MeV]}
& 270.8
& 281.4
& 325.5
& 281.1
& 271.6\\
a_4 {\rm [MeV]}
& 39.97
& 38.68
& 34.05
& 37.26
& 37.41\\
\hline\hline\\
\end{array}
\end{math}

\vspace{10mm}

TABLE II.  Binding energy and charge radius in spherical nuclei
\vspace{1mm}\\
\begin{math}
\begin{array}{lcccccc}
\hline\hline &
\hspace{4mm}{\rm RHAT1} \hspace{4mm} & 
\hspace{4mm}{\rm RHAT2} \hspace{4mm} & 
\hspace{4mm}{\rm RHAT3} \hspace{4mm} & 
\hspace{4mm}{\rm TM}^a   \hspace{4mm} & 
\hspace{4mm}{\rm NL3}   \hspace{4mm} &
\hspace{4mm}{\rm Exp.}  \hspace{4mm} \\
\hline
{\rm ^{16}O}\\
E/A {\rm [MeV]}
& -7.976
& -7.983
& -8.020
& -8.043
& -8.052
& -7.98\\
r_{ch} {\rm [fm]}
& 2.700
& 2.699
& 2.648
& 2.718
& 2.686
& 2.74\\
{\rm ^{40}Ca}\\
E/A {\rm [MeV]}
& -8.561
& -8.557
& -8.517
& -8.503
& -8.547
& -8.55\\
r_{ch} {\rm [fm]}
& 3.431
& 3.434
& 3.398
& 3.513
& 3.437
& 3.45\\
{\rm ^{90}Zr}\\
E/A {\rm [MeV]}
& -8.701
& -8.702
& -8.656
& -8.706
& -8.687
& -8.71\\
r_{ch} {\rm [fm]}
& 4.234
& 4.243
& 4.221
& 4.257
& 4.241
& 4.26\\
{\rm ^{116}Sn}\\
E/A {\rm [MeV]}
& -8.518
& -8.521
& -8.474
& -8.507
& -8.485
& -8.52\\
r_{ch} {\rm [fm]}
& 4.576
& 4.563
& 4.569
& 4.605
& 4.709
& 4.63\\
{\rm ^{208}Pb}\\
E/A {\rm [MeV]}
& -7.859
& -7.877
& -7.870
& -7.875
& -7.879
& -7.87\\
r_{ch} {\rm [fm]}
& 5.489
& 5.498
& 5.481
& 5.489
& 5.494
& 5.50\\
\hline\hline
\end{array}
\end{math}
\begin{description}
\item{$^{a}$}
The TM1 parameter set is supposed to be used above $A>40$, hence we write in 
$^{16}$O and $^{40}$Ca the results with the TM2 parameter set, which is arranged to describe
the small mass nuclei.
\end{description}

\clearpage
\vspace{5mm}

TABLE III.  Energy-weighted sums of 
$B(E\lambda)$ over positive-energy RRPA states in unit of
e$^2$b$^{\lambda}\cdot$MeV. The classical EWSR values (non-relativistic model)
\cite{BM75} are also shown for comparison.
\vspace{1mm}\\
\begin{math}
\begin{array}{lccccccc}
\hline\hline
&\hspace{4mm}{\rm RHAT1} \hspace{4mm} & 
\hspace{4mm}{\rm RHAT2} \hspace{4mm} & 
\hspace{4mm}{\rm RHAT3} \hspace{4mm} & 
\hspace{4mm}{\rm TM1}   \hspace{4mm} & 
\hspace{4mm}{\rm NL3}   \hspace{4mm} &
\hspace{4mm}{\rm Classical}^a  \hspace{4mm} \\
{\rm ^{90}Zr}\\
\hline
E0^{b} & 0.467 & 0.471 & 0.479 & 0.491 & 0.494 & 0.470\\
E1^{c} & 3.572 & 3.587 & 3.655 & 4.368 & 4.224 & 3.303\\
E2     & 6.843 & 6.880 & 6.917 & 7.315 & 7.344 & 5.878\\
\hline
{\rm ^{116}Sn}\\
\hline
E0^{b} & 0.819 & 0.824 & 0.831 & 0.900 & 0.902 & 0.688 \\
E1^{c} & 4.578 & 4.599 & 4.599 & 5.517 & 5.334 & 4.229 \\
E2     & 10.59 & 10.64 & 10.71 & 11.32 & 11.38 & 8.597 \\
\hline
{\rm ^{208}Pb}\\
\hline
E0     & 2.177 & 2.179 & 2.201 & 2.374 & 2.383 & 1.626 \\
E1     & 8.044 & 8.083 & 8.244 & 9.680 & 9.360 & 7.384 \\
E2     & 27.87 & 27.95 & 27.98 & 29.90 & 29.98 & 20.33 \\
\hline\hline
\end{array}
\end{math}
\begin{description}
\item{$^{a}$}
The radial moments $\langle r^{\lambda}\rangle _p$ in the
classical EWSR are calculated with the charge distribution from
the MFT with parameter set RHAT2.\\
\end{description}

\vspace{5mm}

TABLE IV.  ISGQR centroid-energy positions (MeV). 
The contribution from the low-lying discrete states is eliminated in the RHAT1 results.
\vspace{1mm}\\
\begin{math}
\begin{array}{lccccccc}
\hline\hline
&\hspace{8mm}{\rm RHAT1} \hspace{8mm} & 
\hspace{8mm}{\rm Experiment}  \hspace{8mm} \\
\hline
{\rm ^{90}Zr }  & 14.86 & 14.65 \pm 0.20^a\\
{\rm ^{116}Sn } & 13.75 & 13.50 \pm 0.35^b\\
{\rm ^{208}Pb } & 11.39 & 10.89 \pm 0.30^b\\
\hline\hline
\end{array}
\end{math}
\begin{description}
\item{$^{a}$}
Reference \cite{YO041}
\item{$^{b}$}
Reference \cite{YO042}
\end{description}

\clearpage

\end{document}